\documentclass[10pt,conference]{IEEEtran}
\usepackage{cite}
\usepackage{amsmath,amssymb,amsfonts}
\usepackage{algorithm}
\usepackage{graphicx}
\usepackage{textcomp}
\usepackage{xcolor}
\usepackage[hyphens]{url}
\usepackage{fancyhdr}
\usepackage{hyperref}
\usepackage{eso-pic}
\usepackage{booktabs}
\usepackage{tabularx}
\usepackage[libertine]{newtxmath}

\usepackage{xargs}
\usepackage[colorinlistoftodos,prependcaption,textsize=tiny]{todonotes}
\newcommandx{\unsure}[2][1=]{\todo[linecolor=red,backgroundcolor=red!25,bordercolor=red,#1]{#2}}
\newcommandx{\change}[2][1=]{\todo[linecolor=blue,backgroundcolor=blue!25,bordercolor=blue,#1]{#2}}
\newcommandx{\info}[2][1=]{\todo[linecolor=green,backgroundcolor=green!25,bordercolor=green,#1]{#2}}
\newcommandx{\improvement}[2][1=]{\todo[linecolor=Plum,backgroundcolor=Plum!25,bordercolor=Plum,#1]{#2}}
\newcommandx{\thiswillnotshow}[2][1=]{\todo[disable,#1]{#2}}

\usepackage{algpseudocodex}
\usepackage{xspace}

\algrenewcommand\algorithmicindent{1.25em}
\algrenewcommand\alglinenumber[1]{%
  \scriptsize\color{gray}\sffamily #1%
}

\algrenewcommand\algorithmiccomment[1]{%
  \hfill\textcolor{gray}{\(\triangleright\)~#1}%
}

\algnewcommand{\And}{\textbf{and}\xspace}
\renewcommand{\And}{\ensuremath{\DOTSB\;\mathchar"3026 \;\mskip\thickmuskip}}
\algnewcommand{\Is}{\textbf{is}\xspace}

\usepackage{isomath}
\providecommand{\mathsfit}[1]{{#1}}
\newcommand{\var}[1]{\mathsfit{#1}}

\usepackage{circledtext}\circledtextset{resize=real}
\ExplSyntaxOn
\cs_if_exist:NT \draw_set_linewidth:n { \cs_gset_eq:NN \draw_linewidth:n \draw_set_linewidth:n }
\cs_if_exist:NT \draw_set_dash_pattern:nn { \cs_gset_eq:NN \draw_dash_pattern:nn \draw_set_dash_pattern:nn }
\ExplSyntaxOff
\providecommand{\clabel}[1]{}\renewcommand{\clabel}[1]{\circledtext*[height=1.9ex,charshrink=0.65]{#1}}
\providecommand{\cnumber}[1]{}\renewcommand{\cnumber}[1]{\textbf{\circledtext*[height=1.9ex,charshrink=0.7]{#1}}}
\providecommand{\bnumber}[1]{}\renewcommand{\bnumber}[1]{\textbf{\circledtext*[boxtype=OO,height=1.7ex,charshrink=0.8]{#1}}}

\usepackage{pifont}

\newcommand{\hpcayear}{2027}
\newcommand{\hpcasubmissionnumber}{1288}

\newcommand\hpcaauthors{Sangbu Yun\textsuperscript{1}, Jung Gyu Min\textsuperscript{2}, Jin Hyun\textsuperscript{2}, Jeonghwan Ahn\textsuperscript{1}, and Youngjoo Lee\textsuperscript{1}}
\newcommand\hpcaaffiliation{\textsuperscript{1}KAIST, \textsuperscript{2}POSTECH}
\newcommand\hpcaemail{\{boo, jhahn09, youngjoo\}@kaist.ac.kr, \{minjg.epiclab, jinhyun.epiclab\}@gmail.com}
\def\hpcacameraready{}

\title{TwinHash: Two-Sided Object Identity Management\\for Time- and Energy-Efficient Agentic Workflows}

\newcommand{\sysname}{\text{TwinHash}}

\graphicspath{{../figures_current/}{../example/}{figs/300ppi/}{figs/paper_figs/}}

\author{
  \ifdefined\hpcacameraready
    \IEEEauthorblockN{\hpcaauthors{}}
      \IEEEauthorblockA{
        \hpcaaffiliation{} \\
        \hpcaemail{}
      }
  \else
    \IEEEauthorblockN{\normalsize{HPCA \hpcayear{} Submission
      \textbf{\#\hpcasubmissionnumber{}}} \\
      \IEEEauthorblockA{
        Confidential Draft \\
        Do NOT Distribute!!
      }
    }
  \fi 
}

\fancypagestyle{camerareadyfirstpage}{%
  \fancyhead{}
  
  \fancyhead[C]{
    \ifdefined\aeopen
    \parbox[][12mm][t]{13.5cm}{\hpcayear{} IEEE International Symposium on High-Performance Computer Architecture (HPCA)}    
    \else
      \ifdefined\aereviewed
      \parbox[][12mm][t]{13.5cm}{\hpcayear{} IEEE International Symposium on High-Performance Computer Architecture (HPCA)}
      \else
      \ifdefined\aereproduced
      \parbox[][12mm][t]{13.5cm}{\hpcayear{} IEEE International Symposium on High-Performance Computer Architecture (HPCA)}
      \else
      \parbox[][0mm][t]{13.5cm}{\hpcayear{} IEEE International Symposium on High-Performance Computer Architecture (HPCA)}
    \fi 
    \fi 
    \fi 
    \ifdefined\aeopen 
      \includegraphics[width=12mm,height=12mm]{ae-badges/open-research-objects.pdf}
    \fi 
    \ifdefined\aereviewed
      \includegraphics[width=12mm,height=12mm]{ae-badges/research-objects-reviewed.pdf}
    \fi 
    \ifdefined\aereproduced
      \includegraphics[width=12mm,height=12mm]{ae-badges/results-reproduced.pdf}
    \fi
  }
  \fancyfoot[C]{}
}
\begin{document}
\maketitle

\ifdefined\hpcacameraready 
  \thispagestyle{camerareadyfirstpage}
  \pagestyle{empty}
\else
  \thispagestyle{plain}
  \pagestyle{plain}
\fi

\newcommand{\hpcaheight}{0mm}
\ifdefined\eaopen
\renewcommand{\hpcaheight}{12mm}
\fi

\thispagestyle{empty}
\newif\ifieeenotice
\ieeenoticetrue
\ifieeenotice
\AddToShipoutPictureFG*{\put(\LenToUnit{0.5\paperwidth},\LenToUnit{\dimexpr\paperheight-0.28in\relax}){\makebox(0,0)[t]{\parbox[t]{\textwidth}{\centering\scriptsize \copyright\,2027 IEEE. Personal use of this material is permitted. Permission from IEEE must be obtained for all other uses, in any current or future media, including reprinting/republishing this material for advertising or promotional purposes, creating new collective works, for resale or redistribution to servers or lists, or reuse of any copyrighted component of this work in other works.}}}}
\fi

\begin{abstract}

Agentic artificial intelligence (AI) services interleave large language model (LLM) generation with external tool execution, and increasingly rely on search-based workflows that repeatedly revisit shared reasoning states and computations.
These patterns create substantial redundancy across tool outputs, LLM responses, and key-value (KV) prefixes, yet existing serving systems primarily optimize shared prompt prefixes, while naive input-keyed caching is unsafe when identical requests produce different outputs.
We present \sysname{}, a unified object management architecture that assigns each object an input-side hash key to recognize repeated requests and an output-side content fingerprint to validate and locate stored state, complemented by type-specific reuse guards that preserve correctness and response diversity.
A software object identity table maintains reusable objects across requests and trajectories, while a hardware object placement table tracks their residency across hierarchical memory tiers so that evicted state can be retrieved rather than recomputed.
Across seven agent frameworks and five benchmark suites, reuse follows workflow structure more than benchmark identity. \sysname{} serves up to 81\% of tool outputs wherever deterministic tool reads recur, and up to 92\% of LLM responses in search-based frameworks.
Retrieving an evicted KV-prefix from DRAM reduces TTFT by up to 5.6$\times$ on Qwen3-32B, with greater gains projected for larger dense models.
Under search-based frameworks, \sysname{} reduces end-to-end wall-clock time by up to 87\% and measured task energy by up to 81\% over a prefix-cached baseline, while a per-key serving cap controls the quality-reuse trade-off in adaptive search.
These results demonstrate that unified identity, safety, and placement management can convert structural redundancy in emerging agentic workflows into substantial serving-efficiency gains.

\end{abstract}

\section{Introduction}

Nowadays, large language models (LLMs) have emerged as core building blocks for coding, question answering, and knowledge-intensive tasks~\cite{gpt3,gpt4}.
Techniques like chain-of-thought (CoT) and tree-structured prompting elevate single-step generation into multi-step reasoning traces~\cite{cot,tot}.
Furthermore, agentic workflows introduce tool use with an outer action loop where the model issues API calls, executes code, or queries a retriever, conditioning its subsequent steps on the resulting observations~\cite{react,toolformer,reflexion}.
Multi-agent systems further scale these loops by distributing them across role-specialized collaborators~\cite{camel,autogen,chatdev,metagpt}.
Through the progression of agentic workflows, the execution graph of a request is increasingly built at runtime rather than fixed in advance.
Consequently, benchmarks such as AgentBench, SWE-bench, and $\tau$-bench evaluate agents on long-horizon tasks requiring multiple rounds of tool feedback rather than isolated, single-turn prompts~\cite{agentbench,swebench,sweagent,taubench}.

The transition into tool-augmented service offloads computational burden from the model itself onto external tools.
Recent agent frameworks consequently spend a large fraction of their end-to-end wall-clock time waiting on tool execution rather than on model inference~\cite{react,reflexion,lats,llmcompiler}.
Modern frameworks exacerbate this overhead by adaptively widening inference-time tree searches~\cite{abmcts} or searching over entire workflow programs~\cite{aflow}, thereby increasing the algorithmic fan-out that multiplies both tool execution and LLM decoding costs.
Consequently, dynamic and tool-augmented reasoning in agentic workflows substantially inflates energy and infrastructure demands compared to conventional LLM services~\cite{costofdynamic}.

The shift toward dynamic workflows presents a fundamental serving problem beyond mere application-level concerns.
Conventional serving schedulers treat tool calls as opaque external waits, causing tool latency to block the critical path while GPU resources idle~\cite{paste,tokencake,parrot,autellix}.
Along with LLM inference, which dominates the footprint of deployed agent services~\cite{react,llmcompiler,costofdynamic}, growing tool-layer workloads further complicate cost optimization, including latency and energy.

Modern agentic workflows rarely resolve tasks via single-shot trajectories.
Instead, advanced frameworks rely on iterative trials, result verification, and alternative approach exploration~\cite{taubench,sweagent,llmcompiler,lats,abmcts,aflow}.
Meanwhile, agents frequently invoke identical tools and LLM decode requests using identical arguments.
The duplicated executions emerge across retries, parallel search branches, and distinct tasks sharing identical state.
Empirical analysis shows that 31-95\% of tool outputs constitute byte-exact duplicates, and 48-94\% of submitted LLM prompts match previous requests across various agentic frameworks~\cite{taubench,lats,abmcts,aflow}.
The observed redundancy manifests across three distinct computational artifacts, including tool invocations, LLM decode requests, and LLM prefill operations.
Re-executing the redundant operations incurs severe systemic overhead, wasting CPU and GPU resources, network bandwidth, and external API budgets.

Standard caching mechanisms represent the most basic approach to exploit computational redundancy.
However, conventional input-keyed caches fail in agentic workflows for two critical reasons.
First, identical queries submitted to external tools frequently produce divergent outputs.
For instance, database lookups yield strictly deterministic byte sequences, whereas \mbox{31-100\%} of repeated retrieval-augmented queries return different results.
Second, massive KV tensor volumes rapidly force cache evictions from hot-tier HBM.
The eviction triggers the recomputation of massive duplicate objects occupying gigabyte-scale storage.
Conventional systems blindly regenerate evicted data even when identical bytes remain untracked within lower-tier memory subsystems.

Our core insight is an architecture that assigns every object a two-sided hash identity.
The \textit{input-side} key functions as a deterministic execution signature, which exclusively drives the elimination of redundant computations.
The key generation applies uniformly across \textit{objects}, including tool outputs, model responses, and KV-segments.
The \textit{output-side} content fingerprint functions as a physical address key and data integrity validator, derived from an output byte sequence.
The underlying system indexes physical addresses using the content fingerprint, preventing physical replication.
Moreover, the fingerprint encourages memory pressure to demote objects to lower-tier memory instead of outright deletion.
Conversely, the fingerprint verifies the integrity of objects promoted to upper-tier memory, so corrupt or misplaced bytes are not served. Further validity rules prevent stale reuse.

To this end, we present \sysname{}, a unified hash-based object cache designed for dynamic agentic workflows.
\sysname{} registers every materialized tool and LLM output in an identity table keyed by an input-side hash, validates candidate hits via an output-side fingerprint, and tracks object residency across the memory hierarchy to fetch rather than recompute evicted data.
Consequently, the system jointly recovers tool-worker CPU cycles, network bandwidth, API budgets, and GPU compute time. Workflow structure governs the savings, as recurring deterministic tool reads benefit any framework, whereas response reuse requires search-based execution.

This paper makes the following primary contributions:
\begin{itemize}
\item \textbf{A characterization of reusability in agentic workflows.}
We show that potential reuse heavily depends on agent frameworks instead of their tasks\;(benchmarks).
Tool output reuse follows tool-request recurrence and is bounded by tool determinism, whereas LLM response repetition is confined to search-based policies with hit rates of \mbox{14-92\%} while single-trajectory policies reach near zero.

\item \textbf{A safe, content-derived identity mechanism.}
We introduce a two-sided hash architecture \sysname{} that captures structural redundancies including cross-task overlaps invisible to prefix- or session-scoped caches.
The reuse guards prevent cache poisoning and thus misuse of cached objects.
The validity model adds tool versions, per-resource state epochs, and sampled audits. In replays with database and file drift, the model serves no stale object when change signals precede dependent reads.

\item \textbf{Unified resource savings and robustness under memory pressure.}
We demonstrate substantial resource savings in tool outputs, LLM responses, and KV-prefixes using a unified caching abstraction.
The proposed method successfully maintains the reuse hit rate under severe memory pressure thanks to its separated identity tables.
In live paired deployments of search-based frameworks, \sysname{} reduces end-to-end serving latency by up to 87\% and measured task energy by up to 81\%.

\end{itemize}

\begin{figure*}[t]
    \centering
    \includegraphics[width=0.98\linewidth]{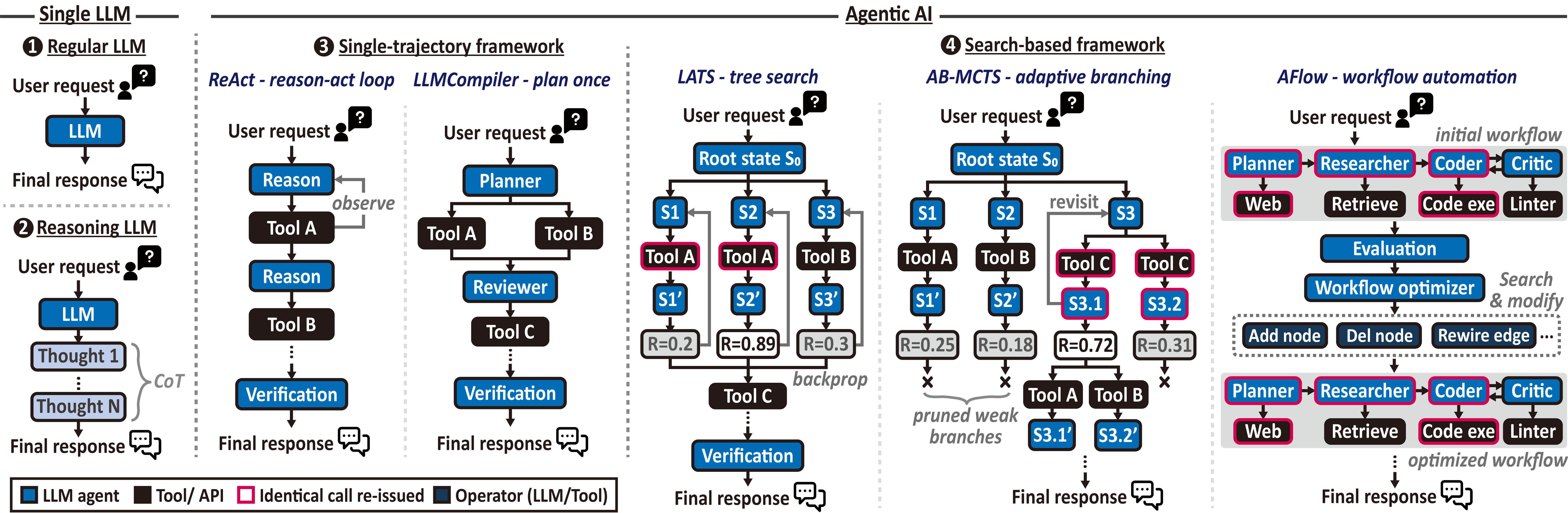}
    \caption{Evolution from single-call LLM serving to agentic workflows.
    Reasoning extends each request into multi-step generation, while tool use introduces external actions; agentic frameworks organize these steps as either a single trajectory or a search over shared states.}
    \label{fig:agentic-frameworks}
\end{figure*}

\section{Background}
\label{sec:background}

\subsection{LLM Serving and Agentic Workflows}
\label{sec:bg-workflows}
LLM inference proceeds in two phases.
The \emph{prefill} phase processes input tokens in parallel and materializes their key and value projections in a KV cache, whereas the \emph{decode} phase autoregressively generates output tokens using the populated cache~\cite{vaswani2017}.
To maximize throughput, serving systems rely on continuous batching and fine-grained KV-cache management, as in vLLM's PagedAttention~\cite{vllm}.
For agentic workloads, this design depends on maintaining a sufficiently large pool of ready-to-run model requests.
External tool invocations suspend individual trajectories and shrink this pool; when the remaining request concurrency is insufficient, batch occupancy decreases and GPU resources become underutilized~\cite{costofdynamic}.

Fig.~\ref{fig:agentic-frameworks} illustrates the architectural evolution that gives rise to this utilization challenge.
Standard LLM deployments expose a simple request-in, text-out interface~\cite{gpt3,gpt4}\;(\cnumber{1}), while reasoning techniques extend each request with a longer internal derivation that trades additional decode compute for improved accuracy~\cite{cot,tot}\;(\cnumber{2}).
Agentic frameworks then place this generation process inside an outer control loop that interleaves model reasoning with external actions~\cite{react,toolformer,reflexion}.
\emph{Single-trajectory} frameworks follow one evolving execution path or commit to a statically planned graph, as in ReAct's reason-act loop and LLMCompiler's plan-once execution~\cite{react,llmcompiler}\;(\cnumber{3}).
\emph{Search-based} frameworks instead explore multiple candidate continuations from shared intermediate states, spanning LATS's tree search, AB-MCTS's adaptive branching, and AFlow's workflow-level optimization~\cite{lats,abmcts,aflow}\;(\cnumber{4}).
Role-specialized multi-agent organization~\cite{camel,autogen,chatdev,metagpt} is orthogonal to this execution-structure axis, as either family can distribute its graph across collaborating agents.

\begin{figure}[t]
\centering
\includegraphics[width=0.99\linewidth]{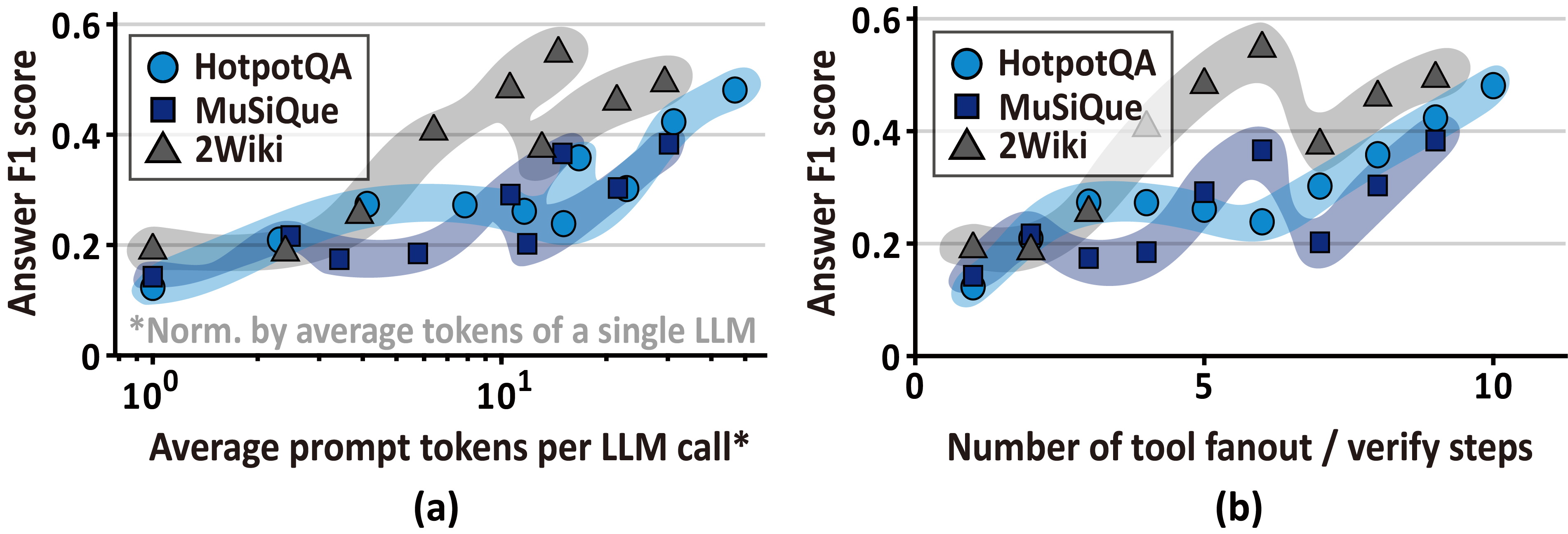}
\caption{Quality versus workflow complexity on multi-hop QA~\cite{hotpotqa,musique,twowikimultihopqa}, measured with Qwen3-14B.
Answer F1 increases with (a) the average prompt tokens per LLM call and (b) the number of tool fan-out and verification steps.}
\label{fig:quality-pressure}
\end{figure}

Transitioning from a naive single-LLM policy to a richer agentic workflow can substantially improve task quality, as demonstrated in Fig.~\ref{fig:quality-pressure}.
These quality gains, however, come with increased structural complexity and thus higher service costs.
For instance, multi-step retrieval-and-verification workflows improve answer F1 scores on multi-hop QA benchmarks from 0.12 to 0.48 on HotpotQA, from 0.14 to 0.38 on MuSiQue, and from 0.19 to 0.49 on 2WikiMultiHopQA~\cite{hotpotqa,musique,twowikimultihopqa}.
Dynamic workflows further determine portions of their execution graph mostly at runtime, introducing irregular model and tool invocations that are difficult to batch efficiently~\cite{costofdynamic}.
The two architectural families expose redundancy in structurally different forms.
Within a single-trajectory framework\;(\cnumber{3}), successive model invocations typically append new observations to a largely shared context, producing substantial KV-prefix overlap even when their complete prompts differ.
Such trajectories may also revisit deterministic tool state as reasoning and verification proceed.
A search-based framework\;(\cnumber{4}) additionally branches from common intermediate states, causing multiple candidates to reuse long context prefixes and repeat identical or equivalent model and tool requests~\cite{got,dylan,lats}.
As inference-time scaling matures, recent systems are increasingly adopting such search-based execution architectures~\cite{abmcts,rstarmath,mctsrag,swesearch,exact,aflow}.

\subsection{Existing Serving Optimizations}

Recently, extensive research has focused on optimizing the LLM-side components of agentic workloads as in Table~\ref{tab:relwork}.
Prefix caches~\cite{vllm,sglang} and KV-offload~\cite{lmcache,mooncake} stacks reuse or relocate prefill state keyed by token position, and workflow-aware methods~\cite{kvflow,parrot} extend the same state by prefetching, overlapping, or scheduling it along predicted execution paths.
Semantic caches instead key on embedding similarity to serve a broader range of requests~\cite{gptcache}.
Cache-augmented generation (CAG)~\cite{cag} exploits predefined knowledge sources to persist state across queries.
Prior work, however, is mostly limited to optimizing the prefill phase of LLM services and agentic workflows.
GPTCache~\cite{gptcache} reuses whole LLM responses for similar requests, yet it prioritizes similarity over exactness and cannot optimize internal flows of agentic services.
Hence, tool invocations and LLM decode responses were not considered as sources of reusable content while preserving accuracy.

\begin{table}[t]
\centering
\caption{Comparison of existing reuse systems\\with how they identify and guard reusable state.}
\label{tab:relwork}
\scriptsize
\setlength{\tabcolsep}{1.5pt}
\resizebox{0.97\linewidth}{!}{%
\begin{tabular}{@{}lllll@{}}
\toprule
System class & Reuses & Keyed by & Cross-req. & Exact \\
\midrule
Prefix cache~\cite{vllm,sglang} & KV prefix & position & prefix & exact \\
KV offload~\cite{lmcache,mooncake} & KV blocks & prefix & prefix & exact \\
Workflow sched.~\cite{kvflow,parrot} & KV, schedule & step graph & program & exact \\
Semantic cache~\cite{gptcache} & responses & similarity & any & none \\
Build cache, CAS~\cite{nix,cloudbuild} & outputs & input / content & any & assumed \\
\textbf{\sysname{}} & tool/resp./KV & input + content & any (tool) & verified \\
\bottomrule
\end{tabular}}
\end{table}

\section{Motivation and Characterization}
\label{sec:motivation}

Targeting efficient agentic service, we motivate \sysname{} through four key observations from various workflows.
First, tool and decode layers represent a substantial yet unmanaged cost\;(\clabel{a}).
Second, tool and decode outputs exhibit significant redundancy that current serving systems fail to reclaim\;(\textbf{\clabel{b}}).
Third, the safety of output reuse depends on a determinism spectrum that naive caches cannot accommodate\;(\clabel{c}).
Finally, a stable identity mechanism alone is insufficient once reusable objects are evicted from the hot memory set\;(\textbf{\clabel{d}}).

\begin{figure}[t]
\centering
\includegraphics[width=1.0\linewidth]{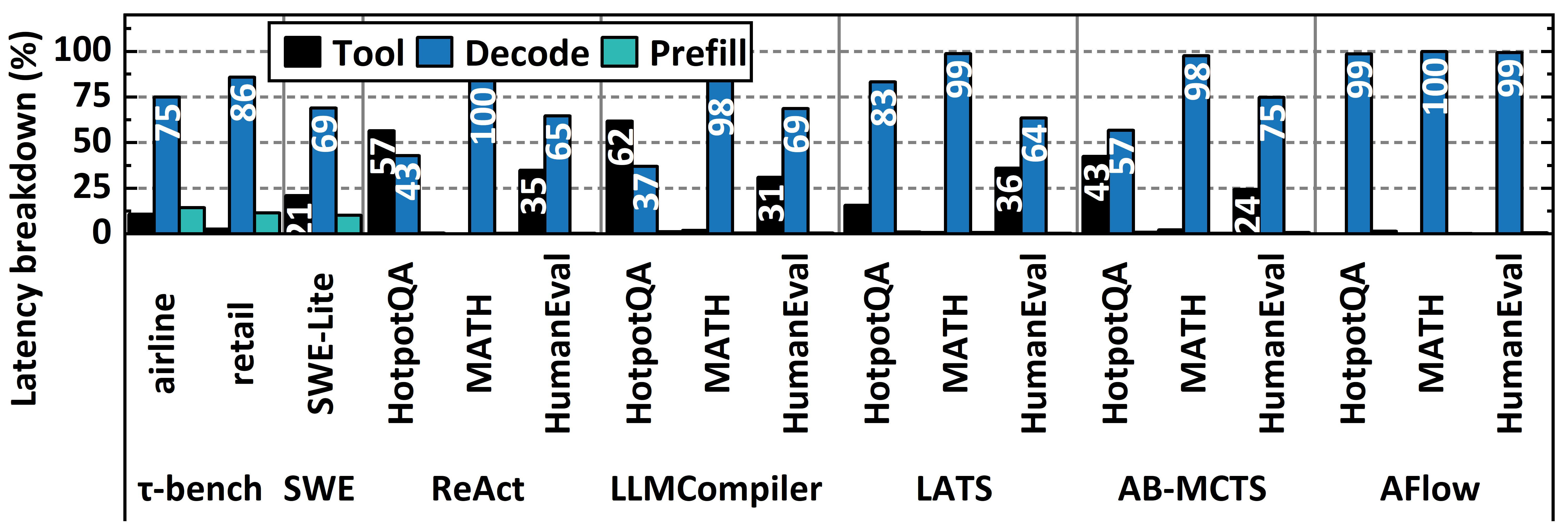}
\caption{Latency breakdown across tool, decode, and prefill for agent frameworks and benchmarks measured on Qwen3-14B with prefix caching~\cite{sglang}.
}
\label{fig:mot-toolshare}
\end{figure}

\begin{figure}[t]
\centering
\includegraphics[width=1.0\linewidth]{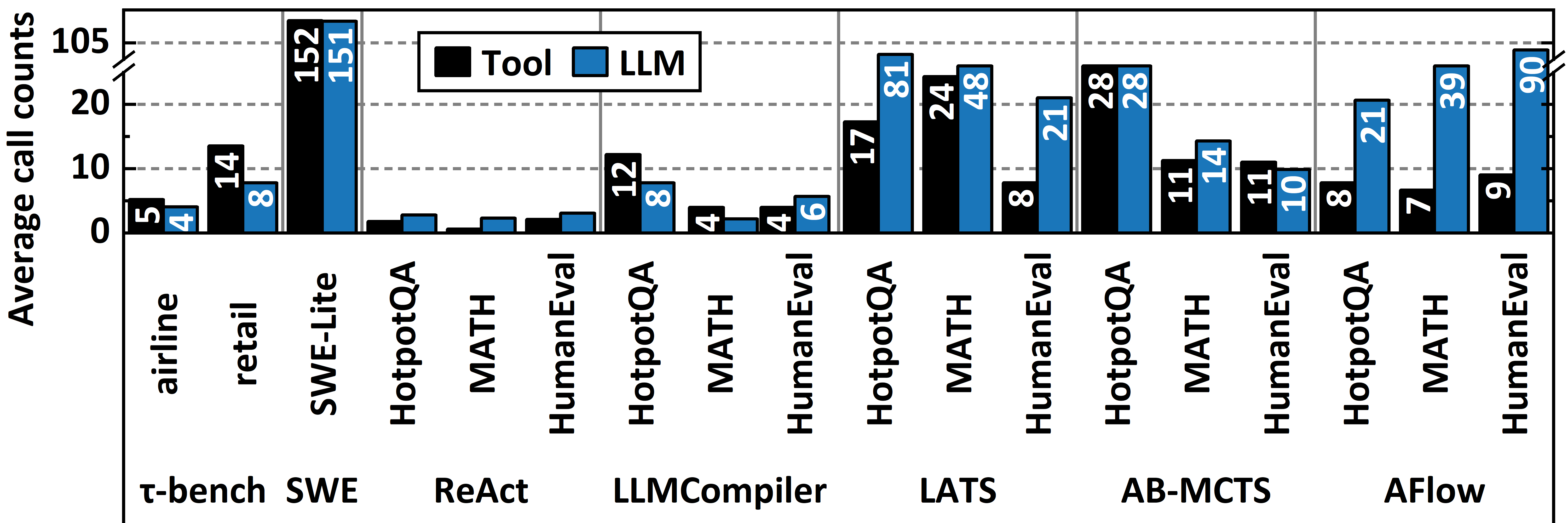}
\caption{Average number of tool and LLM calls per task measured on Qwen3-14B over \mbox{10-20} tasks per experimental cell.
}
\label{fig:mot-toolcount}
\end{figure}

\clabel{a}\;\textbf{Tool and decode layers represent a substantial, unmanaged cost.}
Fig.~\ref{fig:mot-toolshare} shows the latency breakdown between tool, decode, and prefill time across agent frameworks and benchmarks with radix prefix caching~\cite{sglang}.
Tool time spans from issuing a tool call until its result returns, and a verified hit replaces the interval with a lookup.
As agent capabilities expand, agents spend an increasing fraction of their time invoking external tools and LLM decode rather than prefill.
Extensive research on KV-prefix reuse exacerbates the imbalance as the prefill share shrinks further.
On several tool-augmented tasks, tool and decode consume 2.1-61.7\% and 37.0-99.9\% of end-to-end wall-clock time, respectively.
First, tool calls are serialized in conventional schedulers against model generation, which blocks GPU utilization during external execution~\cite{paste,tokencake,autellix,costofdynamic}.
The tool work itself also consumes CPU cycles, network bandwidth, and external API budgets.
Second, LLM decoding consumes a substantial fraction of overall latency due to its autoregressive nature, which drains GPU resources.
These resource demands scale with the call count, and Fig.~\ref{fig:mot-toolcount} shows the profiled agents issuing up to 152 tool and 151 LLM calls per task.
Despite this footprint, current serving stacks fail to treat these outputs as managed state.

\begin{figure}[t]
\centering
\includegraphics[width=1.0\linewidth]{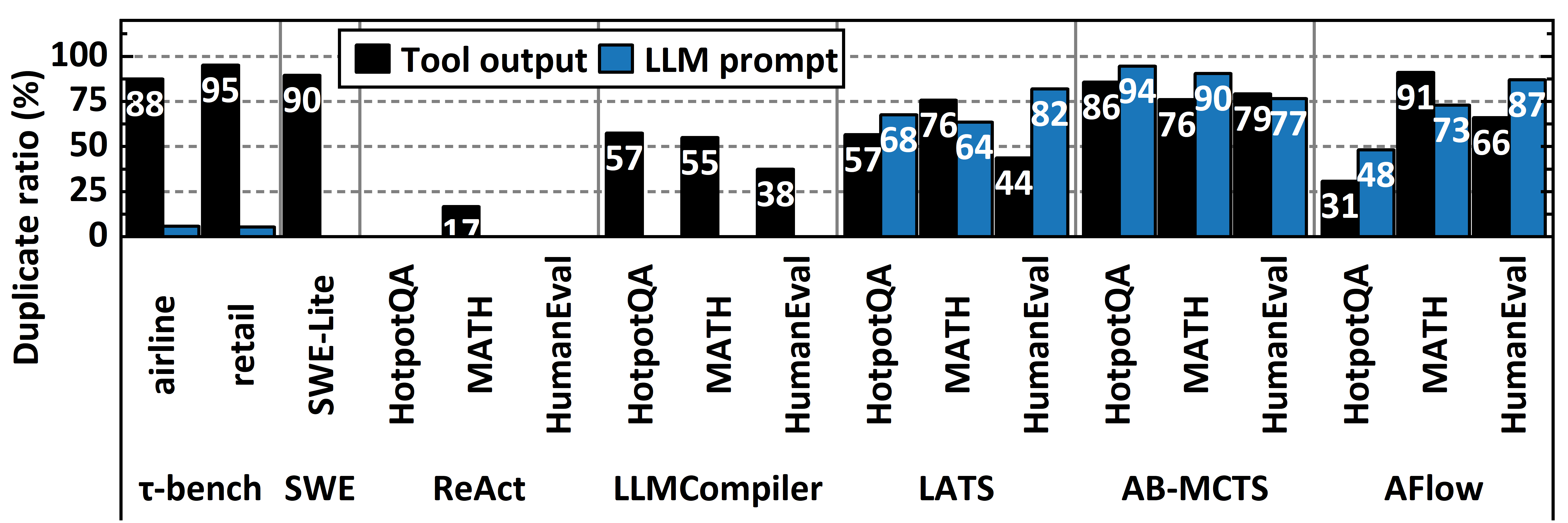}
\caption{Duplicate rates in tool outputs and LLM prompts.
Single-trajectory frameworks ($\tau$-bench, SWE, ReAct, and LLMCompiler) exhibit near-zero LLM-prompt duplication, whereas search-based frameworks (LATS, AB-MCTS, and AFlow) repeat both tool outputs and LLM prompts.
Tool bars count outputs identical to any earlier one (76\% for SWE without empty outputs). LLM bars count requests that repeat except for the seed.}
\label{fig:mot-duplicate}
\end{figure}

\begin{figure}[t]
\centering
\includegraphics[width=1.0\linewidth]{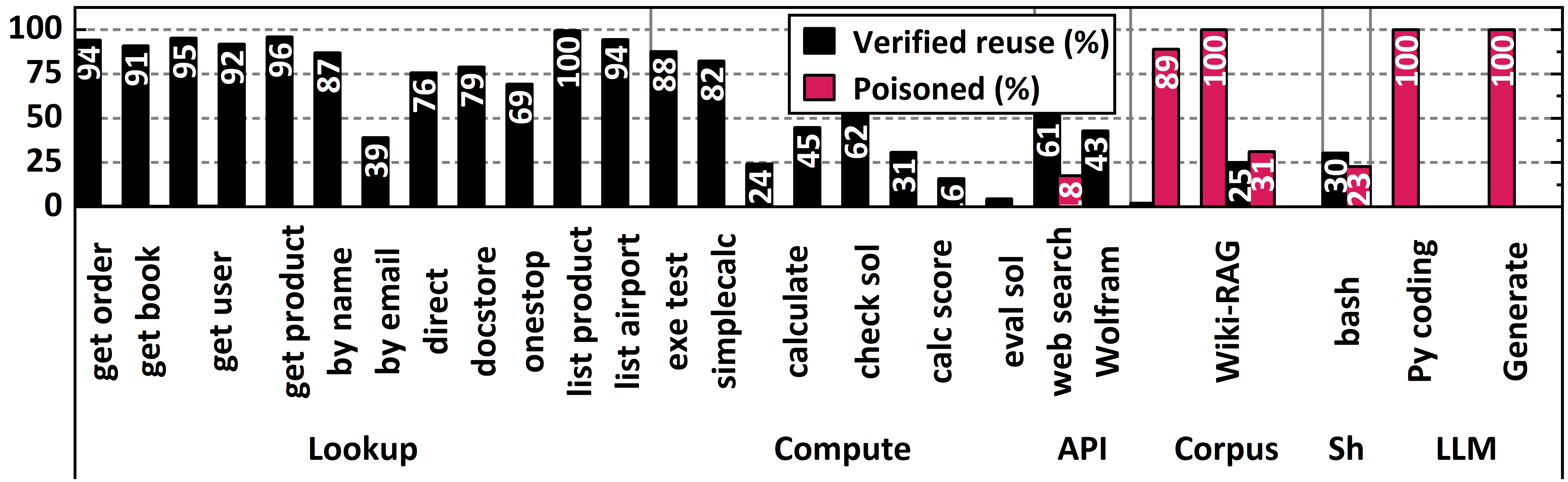}
\caption{Verified reuse and cache-poisoning rates across heterogeneous operators.
Poisoned objects correspond to repeated arguments whose outputs differ and would therefore be served incorrectly by an argument-only cache.}
\label{fig:mot-reuse_safety}
\end{figure}

\begin{figure}[t]
\centering
\includegraphics[width=0.99\linewidth]{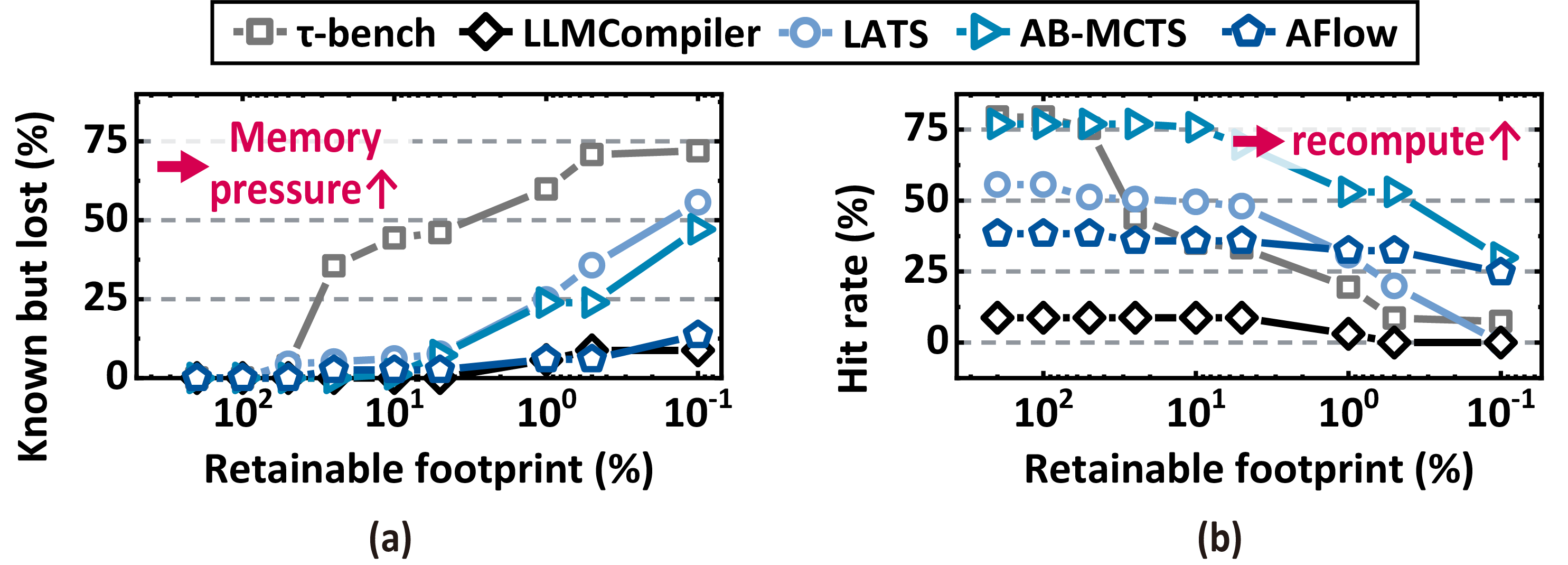}
\caption{Effect of memory pressure on reusable objects.
(a) Known-but-lost fraction and (b) reusable object hit rate as the retainable footprint shrinks.}
\label{fig:mot-pressure}
\end{figure}

\textbf{\clabel{b}\;Tool and decode outputs exhibit massive redundancy.}
Redundancies arise when agents re-issue identical queries, parallel search branches access the same state, or distinct tasks query shared records.
In Fig.~\ref{fig:mot-duplicate}, up to 95.1\% and 94.0\% of tool outputs and LLM prompts were duplicates, respectively.
Tool output repetition appears in every agent framework, whereas LLM prompt repetition emerges mostly under search-based execution.
The main reason for the repetition is structural split and branch expansion that re-issues identical contexts for both tool and LLM.
Re-execution costs much more than serving known objects: a database (DB) query consumes $9.1\times$ more energy than fetching a cached result, and recomputing KV uses $70\times$ more GPU energy than fetching from DRAM.

\begin{figure*}[t]
\centering
\includegraphics[width=0.99\linewidth]{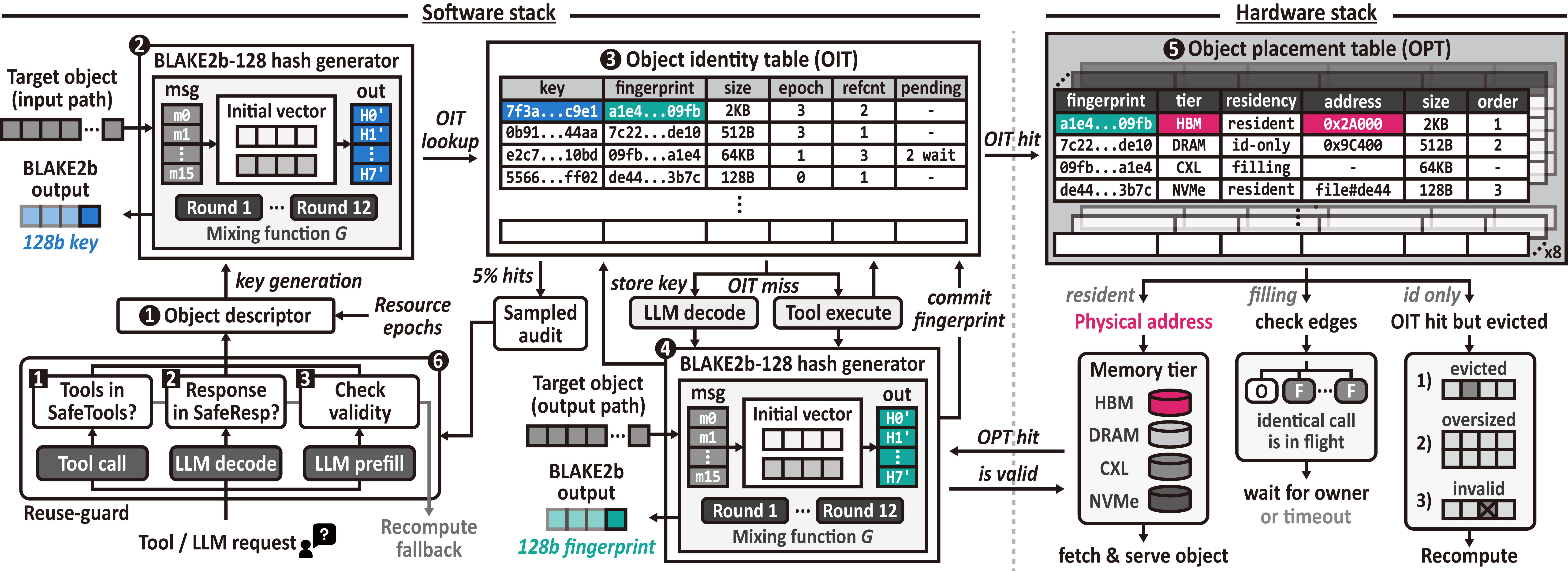}
\caption{Architectural overview of \sysname{}.
Input-side keys identify reusable objects in the OIT, output-side fingerprints validate their content and map physical residency through the OPT, and type-specific reuse guards facilitate safe serving.}
\label{fig:design-overview}
\end{figure*}

\clabel{c}\;\textbf{The safety of reuse follows a determinism spectrum.}
Fig.~\ref{fig:mot-reuse_safety} shows the verified tool and LLM output reuse rate and the error rate of an argument-only cache\;(\textit{poisoned}), which is keyed on the call arguments only without checking the results.
Deterministic DB lookup tools are verified as reusable for 87.9\% of their repeats on average with zero poisoning.
Conversely, retrieval-augmented queries~\cite{rag} and inherently non-deterministic calls including web search, LLM coding and LLM generation suffer poisoning rates of \mbox{18-100\%}.
Moreover, executing the same shell commands across different repositories consistently fails as their environment has changed (shell exec).
Therefore, object reuse within an agentic trajectory requires a solid method to prevent reuse beforehand and verify output content to avoid silent corruption from misuse.

\textbf{\clabel{d}\;Identity mechanisms require placement metadata under memory pressure.}
Identifying a reusable object yields no benefit if its bytes have already been evicted.
Severe memory pressure typically arises during long multi-turn sessions or CAG usage, where huge KV-cache objects quickly exhaust the hot memory tier.
On various benchmarks using Qwen3-14B, Fig.~\ref{fig:mot-pressure} shows that decreasing resident capacity degrades the reusable-object hit rate and forces recomputation.
This pressure intensifies under frontier-scale models such as Llama 3.1\;(405B) and Kimi K3\;(2.8T-A104B)~\cite{llama3,kimik3}, whose weights and KV caches crowd the same memory hierarchy.
Augmenting the object identity with hierarchical placement metadata allows the runtime to fetch evicted data from lower tiers.
Hence, a robust cache should manage content identity with its residency to remain effective beyond the hot set.

These observations set three requirements for \sysname{}.

\noindent \textbf{1)\;Unified caching mechanism for heterogeneous and non-deterministic operators.}
The proposed design must consider two independent features simultaneously:
a unified input-side caching key for heterogeneous objects from either tool or LLM, and reuse safety determined by invocation nature.

\noindent \textbf{2)\;Management for physical address with its identity.}
An object's residency and physical location metadata must remain coupled with its identity, so that the runtime fetches known objects from lower memory tiers rather than recomputing them once the hot tier reaches capacity.

\noindent \textbf{3)\;Reuse policy depends on frameworks rather than tasks.}
The agent framework dictates reuse potential far more than the dataset, as re-issues mostly expose tool and response reuse.
Re-issues that seek diversity from non-determinism should be protected to preserve the framework's service quality.

\section{\sysname{} Architecture Design}
\label{sec:design}

\sysname{} introduces a unified object-management layer at the execution boundaries where an agent's workflow generates reusable objects.
These boundaries include the external tool interface, the model response for structurally repeated prompts, and the KV-segment interface.
Every eligible request follows the same path regardless of type: the runtime canonicalizes the request into a descriptor, digests it into a lookup key, probes a central object directory, and either serves a validated hit or executes natively and commits the fingerprinted output.
A descriptor is the set of structured fields that determine an object's content, listed for each object type in \S\ref{sec:design-identity}.

Fig.~\ref{fig:design-overview} illustrates the \sysname{} unified execution path, which separates object management into two caching data structures.
The software-level object identity table (OIT) tracks the existence of requested objects using hash-generated keys and content fingerprints, while the hardware-level object placement table (OPT) records the physical location of each object across the memory hierarchy.
Since both OIT and OPT are separated from the object bytes, the system identifies objects' existence and residency even after demotion to a lower memory tier.

\subsection{Two-Sided Hash Identity and OIT}
\label{sec:design-identity}

\textbf{The input-side identity mechanism} constructs lookup keys from semantic meaning rather than framework-specific syntax, capturing cross-agent redundancies across heterogeneous reusable objects.
The descriptor fields adapt to each underlying object type\;(\cnumber{1}).
For a \emph{tool invocation}, the descriptor comprises task scope, tool name and version, regularized tool input, and the state epochs of the resources that the call reads, as detailed in~\S\ref{sec:reuse-safety}.
The regularization removes transient runtime fields so that semantically identical calls produce identical keys.
For an \emph{LLM response}, the descriptor incorporates model identifier, prompt messages, and sampling parameters.
Note that the random seed and the request identifier are excluded to prevent fresh seeds from obscuring the structural repetitions of search-based policies.
Sampled diversity is instead managed during the serving phase, as described in~\S\ref{sec:design-serving}.
For a \emph{KV-segment}, the descriptor captures model configuration, exact token IDs, and absolute position range.
Hence, the KV key remains position-bounded unlike tool and response keys.

\begin{figure}[t]
\centering
\includegraphics[width=1.0\linewidth]{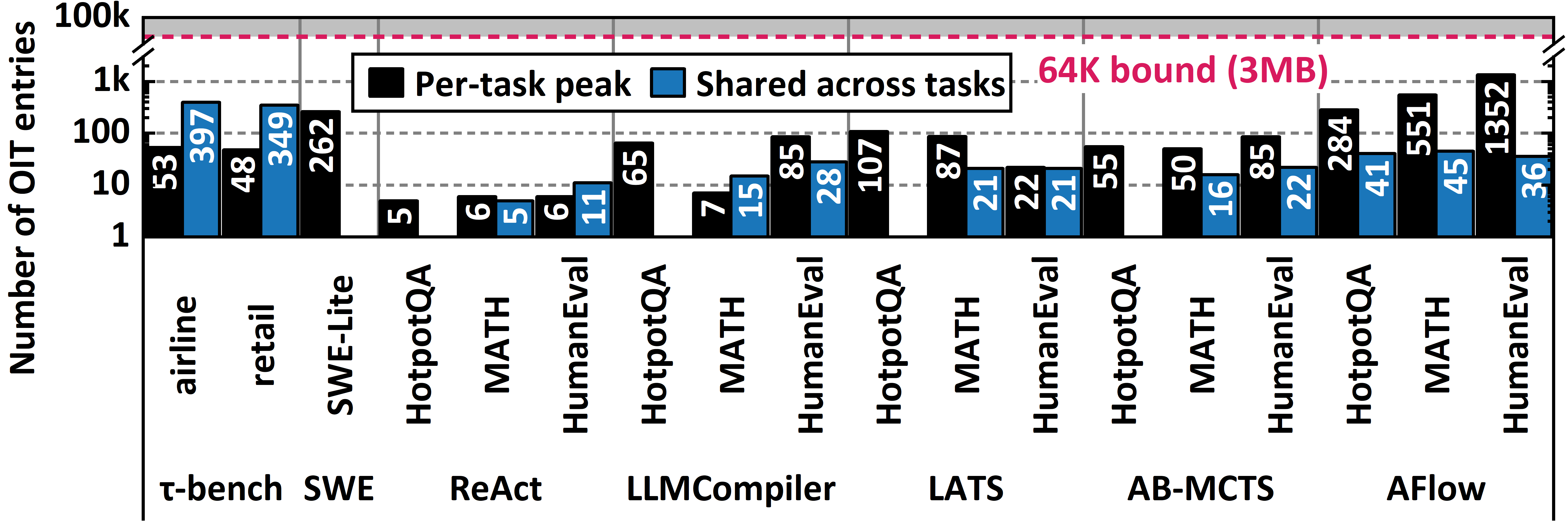}
\caption{OIT entries per cell, given as the per-task peak and the pool shared across tasks, against the default 64K-entry (3\,MB) bound. Cells without exact per-task counts show the whole-run counts, which are an upper bound.}
\label{fig:oit-working-set}
\end{figure}

The runtime hashes each descriptor using 128-bit BLAKE2b and probes a native matching directory\;(\cnumber{2}).
The proposed OIT is designed as a growable hash table in host DRAM and thus has no set conflicts\;(\cnumber{3}).
Per-task identities expire at task end, and shared identities are bounded at 3\,MB (64K entries) under LRU, as in SGLang, vLLM, and LMCache~\cite{sglang,vllm,lmcache}. Hence, an eviction only forgoes a later reuse.
Fig.~\ref{fig:oit-working-set} shows the total number of OIT entries under various workload evaluations, which require at most 1.4K entries.

\textbf{The output-side identity mechanism} manages reuse of the stored object by fingerprint.
When a tool or model is invoked for the first time, the OIT lookup misses, so \sysname{} fingerprints the output with 128-bit BLAKE2b\;(\cnumber{4}) and commits the key and fingerprint to the OIT.
On a hit, \sysname{} retrieves the object's physical address and metadata from the OPT\;(\cnumber{5}).
The fingerprint verifies the retrieved object is valid, as external programs might have mutated the address.
Keys producing an identical fingerprint also collapse into a single copy.

\subsection{Reuse Safety and Reuse-Guard}
\label{sec:reuse-safety}

As analyzed in \S\ref{sec:motivation}, the determinism spectrum for heterogeneous operators, including tools and LLMs, must be considered when classifying the redundancy.
Hence, we introduce an independent structural scoping method \textit{reuse-guard} to manage reuse safety throughout the agentic service\;(\cnumber{6}).

\bnumber{1}\;\textbf{Tool reuse safety.}
Tool reuse becomes unsafe for two distinct reasons: a tool might be inherently non-deterministic, and even an output from a deterministic tool might become outdated or invalid as state changes.
First, \emph{SafeTools} is constructed through fingerprint-based determinism profiling before active reuse.
The profiling pass initially admits candidate read-only tools but executes every invocation natively.
If an identical input-side key produces nonunique fingerprints, the tool is classified as nondeterministic and excluded from SafeTools.
Profiling runs once per tool version, which is part of the key. Between versions, sampled audits re-execute 5\% of the hits of profiled tools off the critical path, and a single mismatch excludes the tool until its next version.
The audits thus detect a wrong declaration within 20 wrong hits on average, without added latency and with at most 4\% more tool calls across the evaluated cells.
Inputs that a tool reads from its own state, such as generated tests, are also included in the key.

Active serving then restricts reuse to the resulting allowlist, while state epochs separately prevent reuse across mutations.
\sysname{} maintains per-resource state epochs, one for each resource a tool declares, such as a database record, a table, a file, or a web page, not one per session. The key includes the epochs of the resources a call reads.
An epoch advances when any session writes the resource or its source signals a change, for example through a change log or a file notification. Any change thus produces a new key as long as the signal precedes dependent reads.
A predicate read depends on its whole table, and a tool that declares no resource depends on a global epoch that every write advances.
Unmodified resources remain at epoch zero and are shared across tasks that start from the same database or repository snapshot, whereas a modified resource is private to the modifying task.
A source without a change signal, such as web search, is admitted to SafeTools without profiling, as its reuse is confined to one task and judged by task accuracy in~\S\ref{sec:reuse-validity}.
The task scope identifies the snapshot for shared resources and the task in all other cases.

\bnumber{2}\;\textbf{Decode response reuse safety.}
Decode response reuse should not rely on output comparison as probabilistic sampling intentionally seeks structural divergence.
Instead, \sysname{} establishes safety via an allowlist \emph{SafeResponses} with three temporal rules.
First, only finalized responses are committed to prevent useless replay of truncated generations.
Second, coalescing concurrent and identical requests is prohibited to preserve mandatory diversity of parallel search algorithms~\cite{lats,abmcts}.
Third, reuse counts are limited for each key to prevent algorithmic insufficiency of the resampling diversity.

\bnumber{3}\;\textbf{KV-prefix reuse safety.}
KV-segment caching structurally guarantees semantic safety through exact-match keying of model configurations and absolute positional encodings.
The primary vulnerability emerges exclusively during cross-tier migrations where logical address bindings occasionally desynchronize from physical bytes.
\sysname{} verifies every promoted segment against its hash-based content fingerprint.

\subsection{Placement and Lifecycle}
\label{sec:design-tables}

The OPT manages overall physical residency of keyed objects across presumed memory subsystems based on \textit{demote-on-eviction} and \textit{promote-on-hit policy}, i.e., tiered-least recently used (LRU).
Like a page table, the OPT is the authoritative map that the runtime's movements consult and update, while the tiering substrate beneath it performs the movements.
Our demotion-on-eviction policy follows a write-back discipline of conventional systems, such as Linux demoting cold pages toward CXL and PMEM nodes during reclaim, and SGLang's hierarchical cache offers the same write-back mode for KV evicted from GPU HBM.
It also departs from offloading methods~\cite{lmcache,mooncake} which copy KV into DRAM proactively at creation and thus consume memory bandwidth for each object by maintaining at least two duplicate KV-segment copies.
Conversely, the demotion prevents duplicates as one fingerprint maps to one physical instance.
Serving a lower-tier hit forces a fetch since a KV-segment must reach the GPU to be consumed; hence, the OPT hit migrates objects toward fast memory.

A hardware OPT is worthwhile for KV-segments, whereas a software directory suffices for the much smaller tool outputs and LLM responses.
Since a framework must locate gigabytes of KV-segments at specific physical tiers, a software-only process can only observe virtual addresses and incur undesired user-space duplicates.
The kernel also migrates pages independently, as in NUMA balancing and memory tiering~\cite{tpp}, so a kernel driver must follow every migration page by page or pin a dedicated pool.
Lookups by the CPU runtime itself need no hardware, since a software table over a pinned pool answers within \mbox{16-58}\,ns.
Lookups by devices such as GPUs, however, reach the software table only through the host CPU and take 4.6\,$\mu$s with a dedicated host core polling at about 3\,W, whereas the OPT answers within 2.25\,ns while dissipating \mbox{3.83-56.4}\,mW.
Moreover, memory pressure from the KV-segments evicts cached tool outputs and LLM responses from HBM, which leads to recompute fallbacks.
Here, the table helps sustain \sysname{}'s hit rate within a session by fetching and promoting the evicted objects from host memory.
The fingerprint thus couples each object identity in the OIT with a single physical copy that the OPT locates in any tier.

As Fig.~\ref{fig:design-overview} shows, the OPT on the host is designed to serve all tiers, and copy engines of every tier update the OPT without the host CPU.
A copy engine updates an entry only after its copy completes, and the DRAM and CXL pools lie outside OS page migration, so a resident entry refers to a complete copy. A referenced entry is not evicted or demoted.

An OPT entry is in one of three residency states to prevent data contamination.
A \emph{resident} object resides in a tier and serves read requests via a direct memory fetch.
A \emph{filling} object is new, with its bytes still arriving, and blocks concurrent readers until completion.
An \emph{identity-only} object retains its OIT presence but has lost its bytes to a bottom-tier eviction or invalidation, so reads fall back to full re-execution of the tool or LLM.
The OPT maintains reference counts so that a writer detaches a private copy before mutating a shared object.

\subsection{Serving Integration}
\label{sec:design-serving}

\begin{algorithm}[t]
\caption{\sysname{} Object Identity Management}
\label{alg:path}
\small
\begin{algorithmic}[1]
\Require Request \(r\) of kind\\
  \(\kappa \in \{\textsc{Tool}, \textsc{Response}, \textsc{KV}\}\) in trajectory \(T\)
\Ensure Output corresponding to \(r\)

\If{\(\kappa = \textsc{Tool}\) \And \(\mathrm{tool}(r) \notin \mathrm{SafeTools}\)}
  \State \(\var{out} \gets \mathrm{ExecuteTool}(r)\)
  \State Advance the epochs of the resources \(r\) writes
  \State \Return \(\var{out}\)
    \Comment{Write path; not reused}
\ElsIf{\(\kappa = \textsc{Response}\) \And \(\mathrm{decode}(r) \notin \mathrm{SafeResponses}\)}
  \State \(\var{out} \gets \mathrm{ExecuteLLM}(r)\)
  \State \Return \(\var{out}\)
    \Comment{Not admitted; not reused}
\EndIf

\State \(\var{key} \gets
  \mathcal{H}\!\left(
    \mathrm{Regularize}
    \left(\mathrm{descriptor}_{\kappa}(r)\right)
  \right)\)
  \Comment{Input-side key}

\If{\(\mathrm{OIT}\) contains a committed entry for \(\var{key}\)}
  \State \(\var{fp} \gets \mathrm{OIT}[\var{key}].\var{fp}\)

  \If{\(\mathrm{OPT}[\var{fp}].\var{state}\) is resident
       \And \(\mathrm{check}_{\kappa}(\var{fp})\)}
    \State \Return \(\mathrm{Fetch}(\mathrm{OPT}[\var{fp}])\)
      \Comment{Promote to the top tier}
  \Else
    \State \textbf{continue}
      \Comment{Identity known, but bytes are lost or unsafe}
  \EndIf
\EndIf

\If{\(\var{key}\) is in flight
     \And \(\mathrm{Produce}_{\kappa}\) is deterministic, non-nested}
  \State Wait for the owner's commit, with a bounded timeout
  \If{the owner commits successfully}
    \State \Return the owner's output
      \Comment{One execution per burst}
  \EndIf
\EndIf

\State Register \(\var{key}\) as in flight
\State \(\var{out} \gets \mathrm{Produce}_{\kappa}(r)\)
  \Comment{Execute, decode, or prefill}

\State \(\var{fp} \gets \mathcal{H}\!\left(\mathrm{Regularize}(\var{out})\right)\)
  \Comment{Output fingerprint using the \(H\)}

\State \(\mathrm{OIT}[\var{key}] \gets \var{fp}\)
\State \(\mathrm{OPT}.\mathrm{insert}(\var{fp}, \var{out})\)
\State Notify followers
\State \Return \(\var{out}\)

\end{algorithmic}
\end{algorithm}

We aim to design the unified management scheme for all eligible requests, including tools and LLMs, as in Algorithm~\ref{alg:path}.
The path contains one admission rule and three object-specific decision points, while all other operations are type-agnostic.
The admission rule originates \textit{reuse-guard}, which manages reuse safety of each object as detailed in~\S\ref{sec:reuse-safety}.

The first object-specific decision is key construction in \S\ref{sec:design-identity}.
As each key includes the state epochs of the resources its call reads, a read issued after a write cannot retrieve an object created before that write.
The second decision validates an entry at serving time.
For tool reuse hits, retrieved bytes must match the fingerprint recorded at insertion.
For response hits, the entry must be finalized and remain within its serving session.
Each response is served at most predefined \(K\) times, and an exceeding request performs a new decode and replaces the cached response to obtain sample diversity.
For KV hits, the segment must match the model configuration and absolute position, and promoted segments are verified with a fingerprint upon arrival.
The third decision determines whether simultaneous requests allow coalescing.
For allowlisted tools, the primary request executes as the owner and concurrent requests wait for its committed result.
Nested tools, which call other cached tools as in LATS on MATH, bypass the wait, since waiting requests could otherwise occupy every worker that the owner's calls need.
In contrast, as search-based frameworks fan out identical prompts to obtain multiple samples, coalescing is prohibited to achieve candidate diversity.

Note that \sysname{} complements existing prefix caching methods rather than replaces them.
Under the validity conditions of \S\ref{sec:reuse-safety}, the reuse guards make serving a cached object safe, and \sysname{} does not mutate a shared prefix.

\subsection{Implementation Costs and Overheads}
\label{sec:design-costs}

The overhead of \sysname{} can be classified as probe latency\;(hashing and lookup), metadata storage\;(OIT), and physical object storage\;(OPT).
In our custom C++/CUDA software kernel of \sysname{}, a directory probe costs 24-32\,$\mu$s per eligible call.
As tool executions consume milliseconds to seconds and LLM calls take seconds to minutes, the kernel call overhead is orders of magnitude below the reuse target.

OIT memory footprints remain negligible because the directory tracks key-fingerprint hashes rather than raw content, where each entry requires only tens of bytes for a 16-byte key digest, a 16-byte content fingerprint, and access counters.
For the OPT, the prototype packs the 16-byte fingerprint, tier tag, physical location, size, reference count, and version fields into a 261-bit payload, roughly 33 bytes.
A 2K-entry OPT managing up to 80\,GB of KV-segments therefore requires only 67.5\,KB of SRAM.
The OPT prototype is 8-way set-associative to mitigate the implementation costs, and a full set demotes its zero-reference tree-PLRU victim.
Implemented in RTL and synthesized using 28-nm CMOS operating at a 444\,MHz clock, the prototype occupies 0.34\,$\mathrm{mm}^2$ while dissipating \mbox{3.83-56.4}\,mW at 0.8\,V.
As lookup and update operate within a cycle, their latency is at most 2.25\,ns.
In RTL simulation over 373,510 operations, including agent traces, the synthesized table matches a reference model on every response at one operation per cycle. The functional model of~\S\ref{sec:methodology} also does not report a hit that the hardware table misses.

\section{Methodology}
\label{sec:methodology}

\textbf{Experimental setup.}
Every experiment runs on a local NVIDIA H200 GPU.
Each primary agent policy uses Qwen3-14B on SGLang with radix prefix caching, a static memory fraction of 0.75, and a page size of 1. Thinking is enabled (8K-token context) for HotpotQA, MATH, and HumanEval and disabled (40K tokens) for $\tau$-bench and mini-SWE.
Qwen2.5-0.5B, Qwen3 dense models up to 32B, Qwen3.6-27B, and the Qwen3-30B-A3B and Qwen3.6-35B-A3B Mixture-of-Experts models serve as KV-fetch anchors on a separate vLLM with LMCache tiering, where we disable the GPU-internal prefix cache to isolate the lower-tier fetch latency of Fig.~\ref{fig:e5-model-scaling}.

We evaluate five representative agent frameworks with two application agents: $\tau$-bench~\cite{taubench}, mini-SWE~\cite{swebench}, ReAct~\cite{react}, LLMCompiler~\cite{llmcompiler}, LATS~\cite{lats}, AB-MCTS~\cite{abmcts}, and AFlow~\cite{aflow}.
Within a unified harness, each policy runs on three benchmarks: HotpotQA~\cite{hotpotqa}, MATH~\cite{math}, and HumanEval~\cite{humaneval}.
Tools are web or document search on HotpotQA, calculators and WolframAlpha on MATH, code or test generation and execution on HumanEval, database reads and writes in $\tau$-bench, and a shell in mini-SWE. AFlow calls only its scorers.
For each framework-benchmark cell, we report the mean over 10 or 15 tasks, with the standard deviation over runs marked.
Fig.~\ref{fig:mot-toolcount} reports the calls per task. An LLM call averages \mbox{0.2K-2.7K} prompt tokens on the three benchmarks, \mbox{4.3K-7.0K} in $\tau$-bench, and 17K in mini-SWE, with 54 to 4.9K output tokens where recorded.
Qwen3-14B reaches \mbox{0.60-1.00} baseline quality in 11 of the 15 framework cells (exact match or F1 on HotpotQA, accuracy on MATH, and pass rate on HumanEval), enough to expose the workflow structure governing reuse.

We further evaluate $\tau$-bench~\cite{taubench} to measure tool-side reuse over vendor model transcripts.
Replaying empirical GPT-4o and Claude-Sonnet transcripts through a function-calling agent under the identity, epoch, and fingerprint rules, we generate 2,809 airline and 8,042 retail tool reads without new model calls.
Since these transcripts lack per-call latency data, we use a local Qwen3-14B run for the latency breakdown in Fig.~\ref{fig:mot-toolshare}.
To probe reuse under mutable repository state, we add a terminal-driven coding agent mini-SWE~\cite{swebench} across 15 SWE-bench Lite instances in isolated containers. \sysname{} reuses only read-only shell commands, such as cat and grep, and every other command advances the epoch.
We further vary three frameworks to isolate how search structure shifts reuse: LATS across per-node samples of 2, 8, and 10, AB-MCTS's adaptive branching on the same tasks, and AFlow's workflow optimizer over 15 validation problems, all on the shared serving stack.

To test whether reuse persists as the hot-tier capacity shrinks, we replay verified event streams against the functional model of our OPT prototype in~\S\ref{sec:design-tables} under capacity-scaled memory tiers, separately from the end-to-end runs of Fig.~\ref{fig:latency_energy_reduction}, with the identity directory unbounded.
Tier service times follow product-class device timings: DDR5 DRAM 0.2$\mu$s,\;25GB/s; CXL 0.6$\mu$s,\;12GB/s; NVMe 80$\mu$s,\;3GB/s.

Across all experiments, we measure end-to-end savings using paired runs that differ only in reuse routing.
The \emph{vanilla} mode is each cell's default setup, the baseline.
The \emph{observe} mode performs all directory lookups to calibrate SafeTools yet bypasses reuse, whereas the \emph{memoize} mode serves verified cache hits.
The vanilla-memoize difference therefore yields \sysname{}'s overall savings, while the observe-memoize difference isolates the effect of reuse.
SGLang enables its radix KV cache identically in all modes, so standard prefix reuse cancels out in the paired ratio, and the reported savings compound on a state-of-the-art prefix-cached baseline.
For decode-layer reuse, measured the same way, a reverse proxy manages the response cache by keying on the prompt and sampling controls, excluding the random seed and refusing to coalesce in-flight model calls.
On the tool side, we instrument the memoizer at AB-MCTS's search tools and AFlow's deterministic scorers.

\textbf{Measurement and metrics.}
We count reuse at two levels for each object type. A \emph{duplicate} in Fig.~\ref{fig:mot-duplicate} is a tool output or LLM request that repeats an earlier one, and a \emph{verified} reuse is a tool call whose request key and output fingerprint both repeat.
Tool hits match the descriptor key, then the output fingerprint. LLM requests match the model, messages, and sampling parameters except the seed, and KV-prefixes match token IDs and absolute positions.
\emph{Verified reuse} in Fig.~\ref{fig:mot-reuse_safety} denotes the fraction of reads whose content check confirms byte-identical output with a previously materialized object.
It therefore captures the repeated reads that are demonstrably safe to reuse and serves as an upper bound on the reuse achievable by a live runtime.
\emph{Response reuse}, by contrast, measures a substitution rate because sampled LLM responses cannot be validated through byte-level content checks before execution, so we audit it
by comparing answer quality between runs with and without substitution.
The tool and LLM time shares in Fig.~\ref{fig:mot-toolshare} are computed by merging overlapping execution intervals.
For tree-search policies executing tool and LLM concurrently, we report the more conservative time share.

\begin{figure}[t]
\centering
\includegraphics[width=1.0\linewidth]{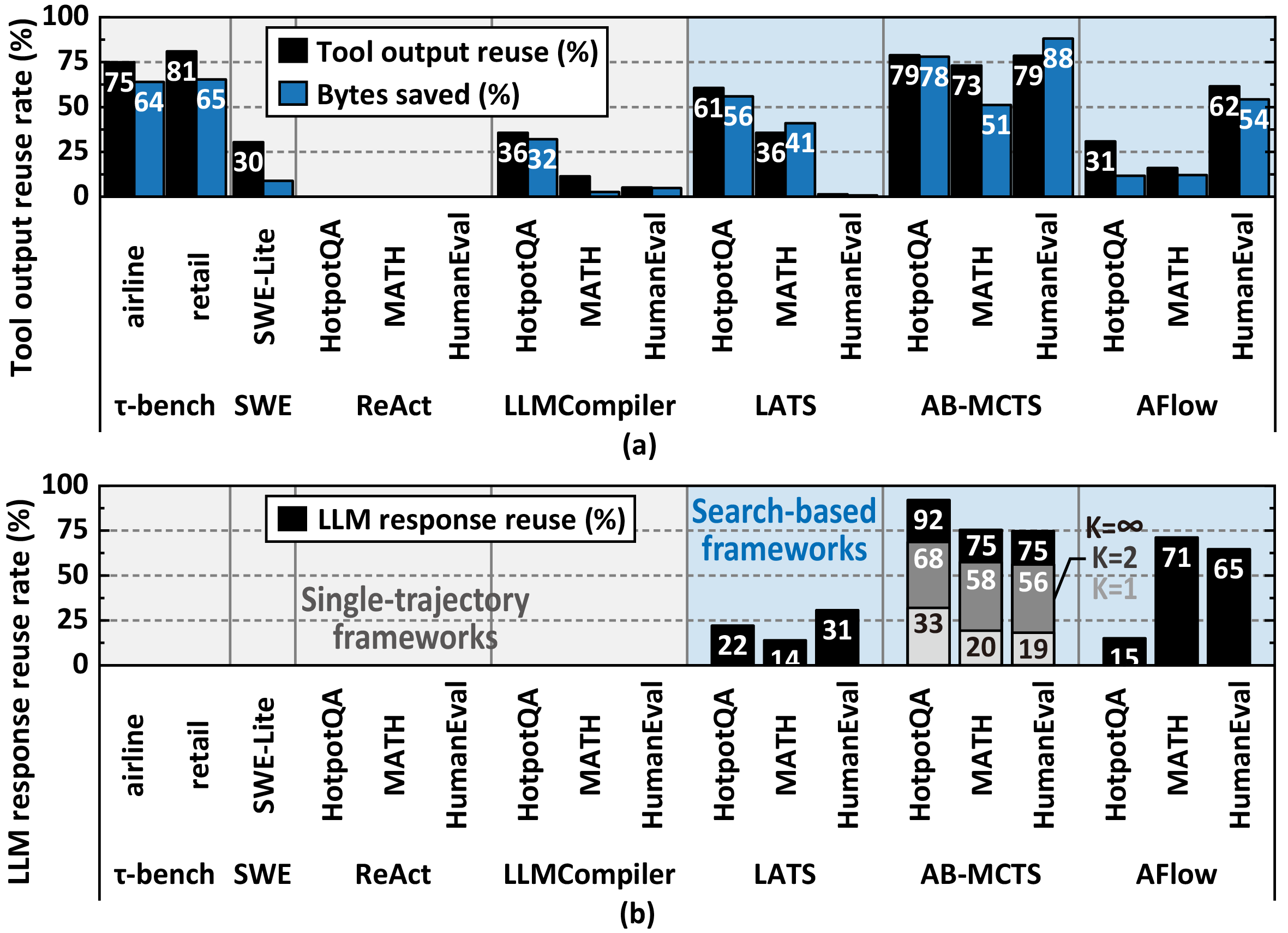}
\caption{Reuse coverage across workloads.
(a) Verified tool output reuse rate and fraction of the output bytes covered, and (b) LLM response reuse rate.}
\label{fig:e1-verified-reuse}
\end{figure}

\mbox{Figs.~\ref{fig:mot-toolshare}-\ref{fig:mot-reuse_safety}} use vanilla or observe runs, except the LLM-prompt duplicates of AB-MCTS and AFlow in Fig.~\ref{fig:mot-duplicate}, counted in the request logs of memoize runs. With \sysname{} off, whole prompts that repeat still form \mbox{81-91\%} of the prefix-cached tokens of AB-MCTS.
Fig.~\ref{fig:e1-verified-reuse}(a) takes verified reuse from replayed calls ($\tau$-bench, mini-SWE, and LATS on HotpotQA), observe runs (AB-MCTS and AFlow), and the cache hits that memoize runs serve after the reuse guards in the other cells. The $\tau$-bench cells replay GPT-4o and Claude-Sonnet transcripts.
Fig.~\ref{fig:e1-verified-reuse}(b) replays recorded LLM requests, and \mbox{Fig.~\ref{fig:e6-placement-gain}(b)-(d)} replays the response logs of memoize runs.
With \sysname{} off, the radix cache serves \mbox{39-99\%} of the prompt tokens, so the baseline already includes KV-prefix reuse.

To evaluate energy at different granularities, we measure per-event energy using idle-corrected RAPL measurements averaged over ten repetitions for each domain.
The reported \emph{measured task energy}, the end-to-end energy consumed while serving one task, combines tool-layer energy with GPU board energy, computed by integrating board power above idle over the corresponding interval of each paired run.
We also characterize energy and latency at the per-hit KV-reuse granularity.
KV-fetch speedup is the ratio of prefill latency to the memory-fetch latency to retrieve the corresponding KV state for a single reuse event.
Unless otherwise specified, all experiments use one request stream and report per-task latency and energy.

\section{Evaluation}
\label{sec:evaluation}

We first analyze reuse per object type, namely external tool outputs, LLM responses, and prefill KV-prefixes, in~\S\ref{sec:reuse_coverage}, then quantify end-to-end latency and energy savings in~\S\ref{sec:e2e_result} and service quality in~\S\ref{sec:service-quality}. Lastly, we evaluate the proposed OPT under severe memory pressure in~\S\ref{sec:place-pressure} and reuse correctness under guards and state drift in~\S\ref{sec:reuse-validity}.

\subsection{Reuse Coverage Across Object Types}
\label{sec:reuse_coverage}

\textbf{Tool output reuse.}
Fig.~\ref{fig:e1-verified-reuse}(a) reports verified reuse, the fraction of tool reads that match a previously materialized object and thus qualify for safe reuse.
The database-backed $\tau$-retail and $\tau$-airline workloads repeatedly issue read-only tool calls with identical arguments, for which the tools produce deterministic outputs.
Consequently, verified reuse reaches 81.0\% and 74.9\%, respectively, accounting for 63.9-65.4\% of their total tool-output bytes.
Cross-task hits alone constitute 34.7\% of the tool-output bytes of both domains and therefore fall outside the scope of prefix- or session-scoped caches.

Search-based frameworks also exhibit substantial tool reuse because tree traversal revisits prior queries and repeatedly invokes deterministic scorers.
AB-MCTS reaches up to 78.9\% verified reuse, covering up to 88.1\% of its tool-output bytes, while LATS and AFlow reach 60.6\% and 61.5\%, respectively.
Single-trajectory agents further show that search structure alone does not determine tool reuse.
The $\tau$-bench function-calling agent achieves high verified reuse because its database-backed workflow repeatedly issues reads with identical arguments.
In contrast, ReAct records 0\% verified reuse in our traces because its linear execution path rarely revisits the same tool inputs.
Tool reuse therefore depends on both argument recurrence and output determinism, as recurring arguments create reuse opportunities within and across sessions, and deterministic outputs make the opportunities safe to exploit.

For the database-backed $\tau$-bench workloads, verified reuse translates directly into reductions in tool-layer energy and execution time.
We measure $869\,\mu$J for a native database read but only $95\,\mu$J for serving its verified result, making native execution $9.1\times$ more energy-intensive.
\sysname{} reduces tool invocation counts by up to 81.0\% on $\tau$-bench and up to 78.9\% in AB-MCTS HotpotQA.
Because these savings arise from bypassing external tool execution rather than GPU inference, they remain independent of the underlying LLM.
\begin{figure}[t]
\centering
\includegraphics[width=0.94\linewidth]{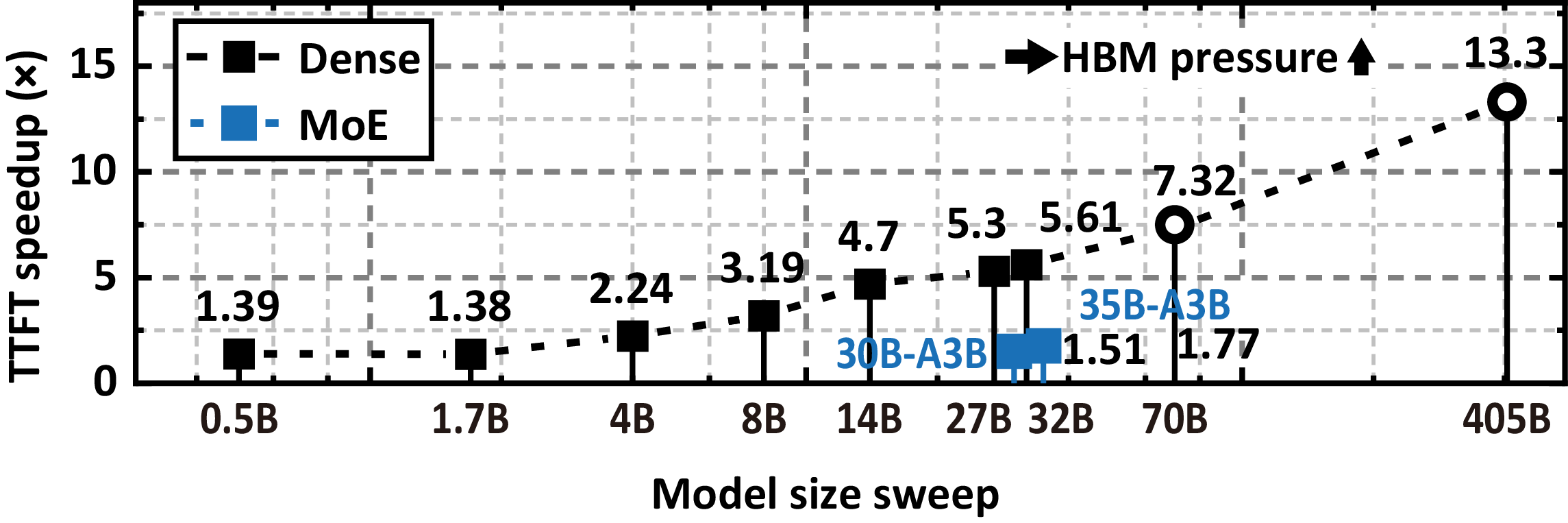}
\caption{TTFT speedup from retrieving an evicted KV-prefix rather than recomputing its prefill state, measured for Qwen-family dense and MoE models up to 35B and modeled for 70B and 405B dense models.}
\label{fig:e5-model-scaling}
\end{figure}
\begin{figure*}[t]
\centering
\includegraphics[width=1.0\linewidth]{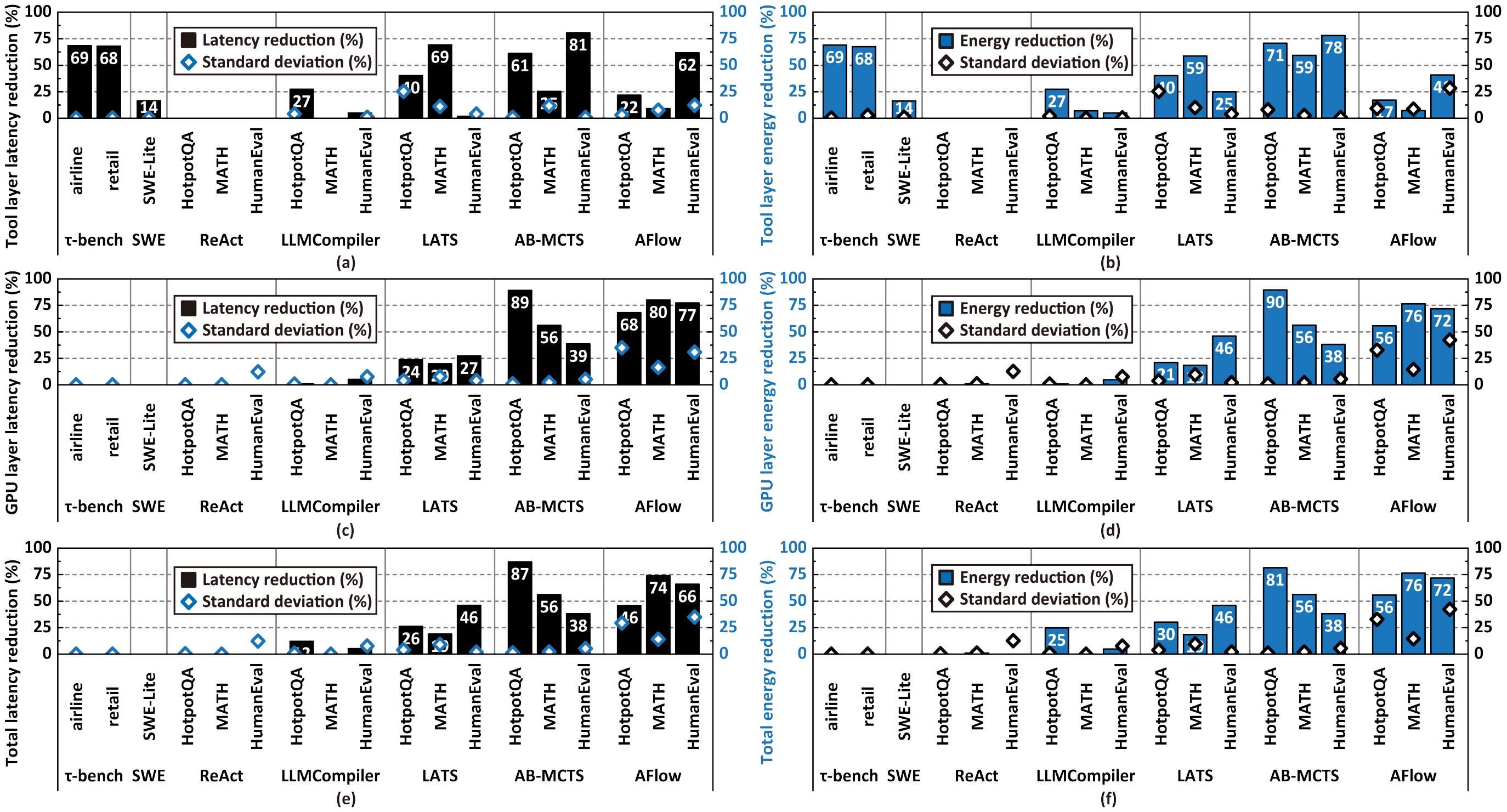}
\caption{Average latency and energy reductions from \sysname{} reuse across agent frameworks and benchmarks, measured on Qwen3-14B in our local serving stack, where the per-key serving cap is set to $K=2$ for AB-MCTS MATH to retain task quality.
(a) Tool-layer latency reduction, (b) tool-layer energy reduction, (c) GPU-layer latency reduction, (d) GPU-layer energy reduction, (e) end-to-end latency reduction, and (f) end-to-end energy reduction.}
\label{fig:latency_energy_reduction}
\end{figure*}

\textbf{Decode-layer response reuse.}
Tool-output reuse avoids external tool execution but leaves LLM decoding, a major part of GPU energy, unchanged.
\emph{Response reuse} targets this remaining cost by serving a completed response when a subsequent model call repeats both the prompt and the sampling configuration.
Fig.~\ref{fig:e1-verified-reuse}(b) shows that the resulting hit rate is strongly governed by how often an agent's execution structure revisits identical model inputs.
With \(K\) of~\S\ref{sec:design-serving} unbounded, leaving only the default limit of 25 hits per key per minute, the cache serves 74.7-92\% of AB-MCTS's model calls, compared with 15.0-71.2\% for AFlow and 13.9-30.7\% for LATS.
These differences reflect how frequently each search strategy revisits completed states: AB-MCTS's adaptive branching produces frequent exact revisits, AFlow's workflow-level optimization yields fewer, and LATS's fixed-width tree expansion exposes the fewest among the evaluated search-based frameworks.

Single-trajectory frameworks exhibit the opposite behavior.
None of the cacheable calls from ReAct, LLMCompiler, and the $\tau$-bench agent repeat within an execution, and none of the coding agent's model calls produce a response hit.
Their monotonically evolving histories and one-shot execution plans rarely re-issue an identical prompt under the same sampling configuration.
Response reuse is therefore concentrated in search-based executions that revisit evaluated model states, rather than simply in technically cacheable model calls.

\textbf{Prefill KV-prefix reuse.}
Beyond tool outputs and decode responses, \sysname{} is designed to apply the same identity and placement mechanism to prefill KV-prefixes, allowing a cache hit to bypass redundant prefill computation.
As KV retrieval or prefill recomputation must complete before decoding, time-to-first-token (TTFT) quantifies the latency benefit of KV-prefix reuse.
Fig.~\ref{fig:e5-model-scaling} reports the TTFT speedup of fetching a DRAM-resident KV-prefix instead of recomputing its prefill state.

For dense models, evaluated using the Qwen3 and Qwen3.6 families, KV retrieval becomes increasingly advantageous as model size grows, reaching a $5.6\times$ TTFT speedup for the 32B model.
The trend arises because, at a fixed prefix length, prefill cost grows with the dense parameter count, whereas the number of transformer layers, KV heads, and cached tokens primarily determines KV-transfer cost.
A scaling model calibrated to our measured results and GQA geometry projects TTFT speedups of $7.3\times$ at 70B and $13.3\times$ at 405B.

The benefit scales less for Mixture-of-Experts (MoE) models from the same families.
While prefill activates only a subset of the parameters, KV retrieval must still transfer cached state across all transformer layers.
Consequently, the evaluated 30B and 35B MoE models achieve a smaller $1.5$-$1.8\times$ TTFT speedup than the $5.6\times$ of the dense 32B model.

\subsection{Latency and Energy Savings}
\label{sec:e2e_result}

Fig.~\ref{fig:latency_energy_reduction} quantifies the latency and energy savings from reuse at the tool, GPU, and end-to-end layers, measured as the paired observe-memoize difference.
As both modes use the same radix prefix cache, all reported gains are measured on top of a state-of-the-art prefix-cached baseline~\cite{sglang}.
Diamonds mark the bootstrap standard deviation over five runs of each cell.

At the tool layer, Fig.~\ref{fig:latency_energy_reduction}(a) and (b) show that latency and energy reductions closely follow the verified tool-reuse rates of \S\ref{sec:reuse_coverage}.
The largest savings arise in workloads with recurring deterministic tool calls, including the database-backed $\tau$-bench and search-based frameworks with repeated scoring, whereas frameworks with little tool-input recurrence obtain correspondingly limited benefit.
At the GPU layer, Fig.~\ref{fig:latency_energy_reduction}(c) and (d) show a sharper separation: savings are concentrated in search-based frameworks because response hits bypass entire decodes.
Single-trajectory agents show essentially negligible GPU-layer reduction because their prompts do not repeat.
Fig.~\ref{fig:latency_energy_reduction}(c) sums the durations of all LLM calls, which overlap under concurrent sampling, rather than wall-clock time.

These layer-specific savings combine into the end-to-end results in Fig.~\ref{fig:latency_energy_reduction}(e) and (f).
AB-MCTS achieves the largest reductions, saving up to 87\% of wall-clock time and 81\% of measured task energy.
AFlow reduces wall-clock time by \mbox{46-74\%} and energy by 56-76\%, while LATS reduces both by \mbox{19-46\%}.
AFlow exhibits the largest variation because its optimizer rewrites the workflow itself between runs, whereas LATS and AB-MCTS resample candidates within a fixed workflow.
Among single-trajectory agents, LLMCompiler on HotpotQA achieves notable end-to-end savings, reducing wall-clock time by 12.1\% and energy by 24.7\%, since web-access API calls cost a significant 1,836\,J of serving energy in total.
ReAct shows task energy savings from \mbox{-3.2\%} to 1.0\%, with bootstrap \mbox{5-95\%} intervals that all include zero. Only 1 of 45 tool outputs in its three cells of Fig.~\ref{fig:mot-duplicate} repeats an earlier one, and \sysname{} adds under 0.02\% to task time.
The $\tau$-bench and mini-SWE agents benefit only at the tool layer, where reuse reduces tool latency and energy by \mbox{68-69\%} and 14\%, respectively. In mini-SWE, most of the saving arises from two of the 15 instances that repeat the same searches.

\subsection{Service Quality and Response Diversity}
\label{sec:service-quality}

\begin{figure}[t]
\centering
\includegraphics[width=1.0\linewidth]{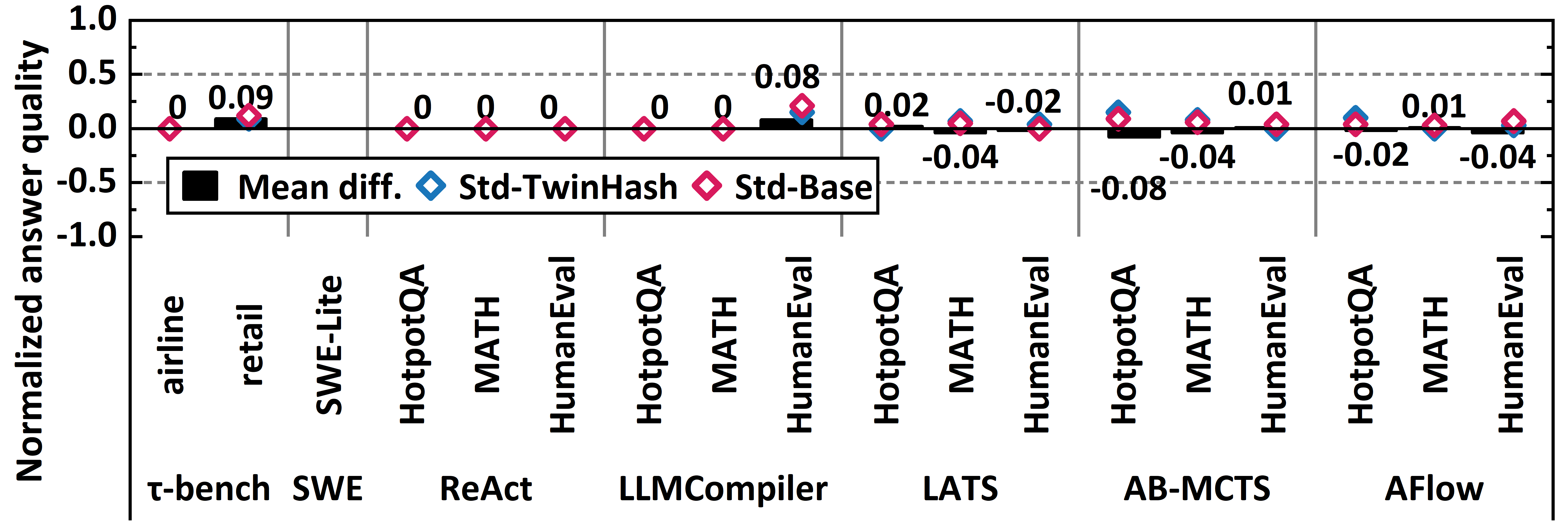}
\caption{Change in normalized task quality under \sysname{} reuse relative to the paired no-reuse configuration across agent frameworks and benchmarks.}
\label{fig:service-quality}
\end{figure}

In Fig.~\ref{fig:service-quality}, service quality remains stable under reuse for most agent-workload pairs.
Greedy-decoding agents, including ReAct and LLMCompiler, exhibit no quality degradation, nor does $\tau$-bench in its stochastic reward metric.
Quality loss arises primarily in adaptive search with stochastic decoding, where repeated sampling of an identical prompt is intentionally used to explore diverse candidate responses.
As discussed in~\S\ref{sec:reuse-safety}, the per-key serving cap can be tuned to preserve the diversity while retaining reuse benefits.
For AB-MCTS on MATH, setting \(K\)$=$\(2\) reduces the quality loss from 0.25 to 0.04 while still reducing time and energy by \(56.1\%\) and \(56.4\%\), compared with \(84.1\%\) and \(84.6\%\) without a cap.
Although \(K\)$=$\(1\) achieves the same measured accuracy, it reduces time and energy by only \(18.2\%\) and \(19.0\%\).
The number of forced fresh decodes per run increases from \(5\) to \(39\) and then to \(63\) as \(K\) tightens.

\subsection{Placement Under Pressure}
\label{sec:place-pressure}
\begin{figure}[t]
\centering
\includegraphics[width=1.0\linewidth]{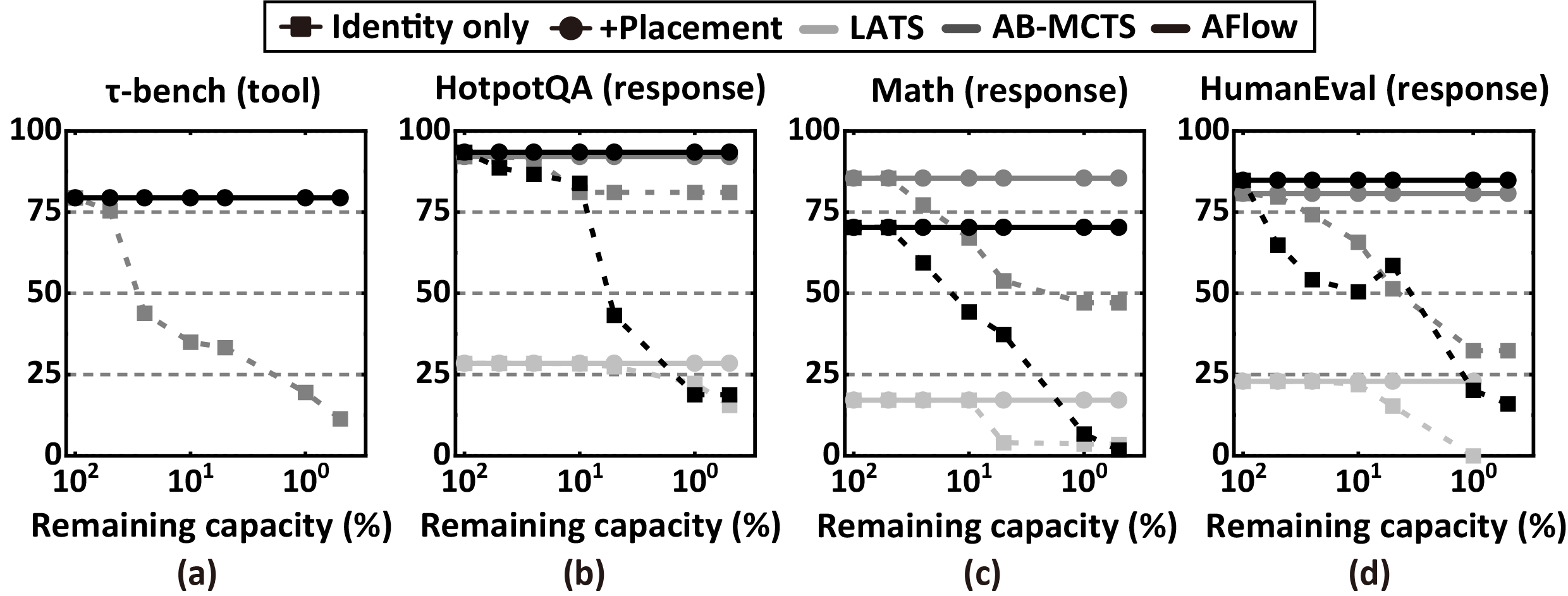}
\caption{Object hit rate under increasing hot-tier memory pressure for (a) $\tau$-bench tool objects and (b)-(d) HotpotQA, MATH, and HumanEval response objects.
Identity-only serving loses reusable state as HBM capacity shrinks, whereas identity-plus-placement sustains each policy's reuse ceiling.}
\label{fig:e6-placement-gain}
\end{figure}

KV-segments dominate hot-tier capacity, but evicting much smaller tool outputs and model responses can eliminate disproportionately valuable reuse.
To isolate the benefit of placement, Fig.~\ref{fig:e6-placement-gain} compares identity-only serving, where an OIT hit without resident bytes triggers re-execution, against identity-plus-placement, where the OPT retrieves the object from a lower tier.
The functional model of our OPT prototype in~\S\ref{sec:methodology} produces the results with presumed product-class tier timings.
As an OPT entry records only a location and a size, the OPT handles tool and response objects like KV-segments.

Hierarchical placement becomes increasingly valuable as the reusable working set exceeds hot-tier capacity.
For $\tau$-bench tool objects, Fig.~\ref{fig:e6-placement-gain}(a) shows the identity-only hit rate falling to 33.3\% at 5\% remaining capacity and 19.5\% at 1\%, whereas identity-plus-placement sustains the 79.4\% verified-reuse ceiling throughout.
In Fig.~\ref{fig:e6-placement-gain}(b)-(d), identity-only serving also progressively loses reusable response objects, while placement preserves each framework's response-reuse ceiling.
The benefit is largest for search-based policies with high structural reuse, where eviction otherwise destroys a larger fraction of reusable computation.
On $\tau$-bench, placement-driven energy savings increase from 3.3\% at $2\times$ pressure to 59.2\% at $200\times$.

Live measurements further validate the modeled tier hierarchy.
Fetching an evicted KV-prefix from CPU DRAM outpaces prefill recomputation by $1.8$-$6.0\times$ in TTFT across 0.8K-6.4K-token prompts on an 8B model, supporting DRAM as an effective lower-tier reuse target, whereas disk retrieval achieves $0.86$-$1.3\times$, indicating that the bottom tier is more valuable for high-capacity retention than for latency-optimized serving.
As severe memory pressure is primarily driven by growing KV state during extended sessions, the value of placement further increases as model and context sizes scale~\cite{cag,llama4,deepseekv3}.

\subsection{\texorpdfstring{Reuse Guard Ablation and State Drift}{Reuse Guard Ablation and State Drift}}
\label{sec:reuse-validity}

\begin{figure}[t]
\centering
\includegraphics[width=0.97\linewidth]{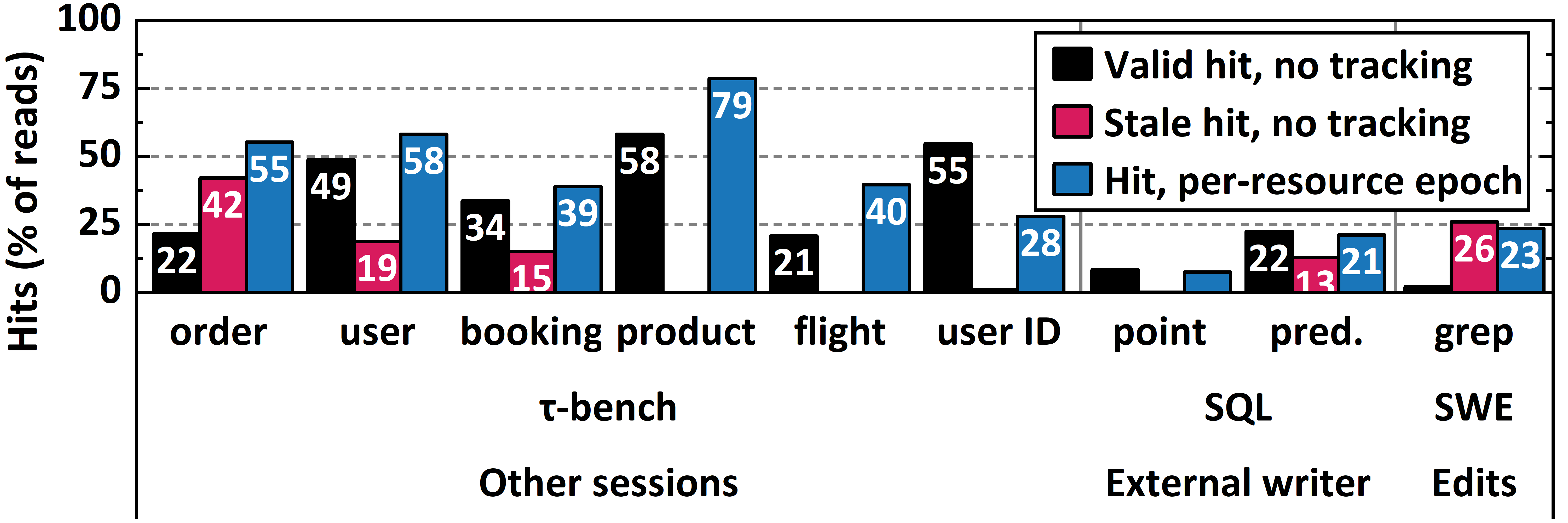}
\caption{Hits per tool type of replayed calls while other sessions, an external writer, or out-of-band edits change the state, with stale hits in red.}
\label{fig:drift}
\end{figure}

Removing one reuse-guard rule at a time shows what each rule prevents. Without output checks, an argument-only cache serves wrong bytes on 251 of 10,330 hits. Without task scopes, shell reads shared across SWE-bench repositories are wrong on all 126 hits, and without state epochs, 22 of 103 reads repeated across a write differ. With all rules, the cache serves no wrong hit.
For responses, without the default limit of~\S\ref{sec:reuse_coverage}, 5 of 9 LATS runs stalled on a single key, and coalescing sampled calls collapsed each 5-way fan-out into one decode.
Our prototype implements the validity rules at low cost, as task-local scope changes only 4 of 1,695 HotpotQA reuse decisions and 5\% audits find no mismatch while re-executing up to 4\% of tool calls. Including audit energy would lower the savings of Fig.~\ref{fig:latency_energy_reduction}(b) by at most 4 points.

Fig.~\ref{fig:drift} evaluates whether per-resource state epochs prevent stale hits in three replays with state drift. With every change signaled before dependent reads, the epochs serve no stale hit and still hit \mbox{7-79\%} of reads.
In $\tau$-bench, 100 airline and 230 retail sessions share one database per domain (\mbox{21-24\%} of tool calls write). The SQL of 4,495 DBBench tasks~\cite{agentbench} queries 152 databases, each shared by eight interleaved sessions. An external writer changes one row of a random table after each statement with probability 0.5. In mini-SWE, 114 out-of-band edits per run each insert a commented-out copy of a line that is read later.
The \emph{no tracking} baseline sees only a session's own writes ($\tau$-bench), writes through the cache (SQL), or the agent's own commands (mini-SWE), and up to 66\% of its database hits and 93\% of its file hits are stale.
Compared with a per-session epoch, per-resource epochs preserve sharing after a session's first write (58\% to 79\% for product lookups), but any change to a scanned record invalidates a table scan (55\% to 28\% for user-ID lookups).
In Fig.~\ref{fig:e1-verified-reuse}(a), $\tau$-bench shares only reads before a session's first write, and per-resource epochs would raise both rates to \mbox{83-88\%} without a stale hit.

Conversely, web results change without a signal. Over 65 days, 28.6\% of queries returned different results, whereas 97\% of repeated searches within a task returned identical results.
LLMCompiler on HotpotQA, which reuses only web search, still retains a task quality of 0.60 with and without reuse.
\section{Discussion and Limitations}
\label{sec:discussion}

\textbf{LLM response reuse and sampling diversity.}
On the AB-MCTS MATH benchmark, unbounded reuse of stored samples under stochastic decoding lowers accuracy from \(0.90\) to \(0.65\).
As~\S\ref{sec:service-quality} shows, a tighter cap than \(K\)$=$\(2\) sacrifices most of the efficiency gain with little additional quality benefit, and each fresh decode replaces the stored sample and restores sampling diversity.
This degradation is thus specific to stochastic decoding, where repeated sampling contributes to exploration.

\textbf{Time and energy crossovers for KV fetching.}
KV-segment reuse exhibits different time and energy crossovers at the NVMe tier.
For Qwen3-8B, an NVMe-tier fetch is faster for the \(848\)-token prompt~(\(1.31\times\)), and recomputation is faster beyond roughly 1.6K tokens, with recompute-to-fetch time ratios of \(0.95\times\) at \(1{,}645\), \(0.99\times\) at \(3{,}244\), and \(0.86\times\) at \(6{,}447\) tokens.
A cold prefill consumes \(72.45\)\,J, whereas a tier hit consumes \(1.04\)\,J.
Recomputation still draws \mbox{\(300\)-\(400\,\mathrm{W}\)} of net GPU board power against an estimated \mbox{\(10\)-\(15\,\mathrm{W}\)} for an NVMe fetch, so the fetch uses an order of magnitude less energy even where recomputation is faster.
The latency crossover thus lies between the modeled CXL tier and NVMe, and the energy crossover beyond NVMe. Increasing model size shifts the latency crossover further in favor of fetching, as prefill cost scales more rapidly with model size than KV transfer.

\textbf{Benefits beyond search-based execution.}
Single-trajectory agents repeat no response, yet tool reuse still saves up to 12.1\% of their time and 24.7\% of their energy, and the saving grows with tool requests that recur across tasks and users.
Evicted KV-prefix retrieval applies to every framework, as the single-trajectory agents of~\S\ref{sec:bg-workflows} also append to a shared context, and the prefix cache serves \mbox{52-99\%} of their prompt tokens.

\textbf{Scaling beyond a single node.}
A cluster can shard the OIT by key hash, with each owner shard storing the in-flight state of its keys. Each node's OPT treats peer memory as a lower tier, as in distributed KV pools~\cite{mooncake}.
In a trace-driven $\tau$-bench simulation, task-affine routing over eight node-local directories retains 96\% of the reuse, and key sharding loses no reuse.

\textbf{Limitations.}
Per-resource state epochs assume that change signals precede dependent reads. Sources without a change signal, such as web search, are reused only within a task.
We validate the OPT prototype through synthesis and trace-driven simulation, while silicon or FPGA integration, physical CXL memory, and multi-node deployment remain future work.

\section{Related Work}
\label{sec:related-work}

\textbf{Agentic frameworks and inference-time scaling.}
CoT~\cite{cot} introduces intermediate reasoning steps, while ToT~\cite{tot} and GoT~\cite{got} organize them into tree- and graph-structured representations.
ReAct~\cite{react} interleaves reasoning with actions and observations, LLMCompiler~\cite{llmcompiler} generates dependency-aware function-call plans, and multi-agent frameworks~\cite{camel,autogen,chatdev,metagpt} coordinate role-specialized collaborators.
Search-based systems~\cite{lats,abmcts,rstarmath,mctsrag,swesearch,exact} further explore multiple reasoning or execution paths, repeatedly invoking tools and models.
\sysname{} complements these frameworks without modifying reasoning or search policies, instead reclaiming repeated work exposed by their execution structures.

\textbf{Agent-aware scheduling.}
Parrot~\cite{parrot} and Autellix~\cite{autellix} schedule agent programs to reduce queueing delays and context churn.
KVFlow~\cite{kvflow} prefetches predicted KV state, FlashAgents~\cite{flashagents} overlaps decoding with prefilling, and PASTE~\cite{paste} and TokenCake~\cite{tokencake} mitigate tool-induced stalls through speculation and wait-aware KV-cache management.
These techniques reduce latency or memory pressure through scheduling, prefetching, speculation, and overlap, whereas \sysname{} avoids verified duplicate execution.

\textbf{Model-side serving reuse.}
vLLM~\cite{vllm}, SGLang~\cite{sglang}, Prompt Cache~\cite{promptcache}, and CAG~\cite{cag} reuse compatible prefixes and position-aware or preloaded model state, while CacheBlend~\cite{cacheblend} extends KV reuse to retrieval-augmented inputs~\cite{rag}.
Other systems move model parameters or KV state across memory tiers, hosts, and storage~\cite{flexgen,llminflash,cachegen,infinigen,mooncake,memserve,lmcache}.
In contrast, \sysname{} generalizes reuse beyond model-native state to tool outputs, model responses, and KV segments through a unified identity abstraction.

\textbf{Content-addressed storage and build caches.}
Content-addressed stores identify data by a hash of its bytes~\cite{venti,lbfs,datadomain}, and copy-on-write file systems share blocks until a write~\cite{btrfs}, as the OPT reference counts do. Build caches key outputs by a hash of their inputs~\cite{nix,nectar,cloudbuild,alacarte}, as does memoization~\cite{michie}.
Input-keyed caches assume hermetic producers, which agentic tools and sampled decoding are not. \sysname{} extends both hashes with versions, state epochs, audits, and serving caps as rules for valid reuse.

\section{Conclusion}

As agentic AI shifts from isolated model calls to dynamically constructed workflows, serving efficiency increasingly depends on eliminating repeated work rather than merely accelerating each invocation.
\sysname{} addresses this opportunity through a unified object-management architecture that spans tool outputs, LLM responses, and KV prefixes using input-side identities, output-side fingerprints, type-specific reuse guards, and hierarchical placement metadata.
Across seven agent frameworks and five benchmark suites, we find that reuse is governed primarily by workflow structure rather than benchmark identity, with \sysname{} serving up to 81\% of tool outputs and 92\% of LLM responses.
By independently eliminating redundant work at the tool and GPU layers, \sysname{} converts these reuse opportunities into end-to-end reductions of up to 87\% in wall-clock time and 81\% in measured task energy over a prefix-cached baseline.
Extending the same identity to object placement further preserves reusable state under severe memory pressure, increasing the object hit rate from 19.5\% to 79.4\% at 1\% resident capacity.
The hardware OPT records object placement across all tiers without the host CPU, within 2.25\,ns at \mbox{3.83-56.4}\,mW.
Therefore, TwinHash unifies object management across heterogeneous objects, establishing a cost-effective and scalable framework for the emerging era of complex agentic workflows.

\bibliographystyle{IEEEtranS}
\bibliography{refs}

@article{gpt4,
  title        = {{GPT-4} Technical Report},
  author       = {{OpenAI}},
  journal      = {arXiv preprint arXiv:2303.08774},
  year         = {2023}
}

@inproceedings{gpt3,
  title        = {Language Models are Few-Shot Learners},
  author       = {Brown, Tom B. and Mann, Benjamin and Ryder, Nick and Subbiah, Melanie and Kaplan, Jared and Dhariwal, Prafulla and Neelakantan, Arvind and Shyam, Pranav and Sastry, Girish and Askell, Amanda and Agarwal, Sandhini and Herbert-Voss, Ariel and Krueger, Gretchen and Henighan, Tom and Child, Rewon and Ramesh, Aditya and Ziegler, Daniel M. and Wu, Jeffrey and Winter, Clemens and Hesse, Christopher and Chen, Mark and Sigler, Eric and Litwin, Mateusz and Gray, Scott and Chess, Benjamin and Clark, Jack and Berner, Christopher and McCandlish, Sam and Radford, Alec and Sutskever, Ilya and Amodei, Dario},
  booktitle    = {Advances in Neural Information Processing Systems},
  pages        = {1877-1901},
  year         = {2020}
}

@inproceedings{cot,
  title        = {Chain-of-Thought Prompting Elicits Reasoning in Large Language Models},
  author       = {Wei, Jason and Wang, Xuezhi and Schuurmans, Dale and Bosma, Maarten and Ichter, Brian and Xia, Fei and Chi, Ed H. and Le, Quoc V. and Zhou, Denny},
  booktitle    = {Advances in Neural Information Processing Systems},
  year         = {2022}
}

@inproceedings{tot,
  title        = {Tree of Thoughts: Deliberate Problem Solving with Large Language Models},
  author       = {Yao, Shunyu and Yu, Dian and Zhao, Jeffrey and Shafran, Izhak and Griffiths, Thomas L. and Cao, Yuan and Narasimhan, Karthik},
  booktitle    = {Advances in Neural Information Processing Systems},
  year         = {2023}
}

@inproceedings{react,
  title        = {{ReAct}: Synergizing Reasoning and Acting in Language Models},
  author       = {Yao, Shunyu and Zhao, Jeffrey and Yu, Dian and Du, Nan and Shafran, Izhak and Narasimhan, Karthik and Cao, Yuan},
  booktitle    = {International Conference on Learning Representations},
  year         = {2023}
}

@inproceedings{toolformer,
  title        = {Toolformer: Language Models Can Teach Themselves to Use Tools},
  author       = {Schick, Timo and Dwivedi-Yu, Jane and Dess{\`i}, Roberto and Raileanu, Roberta and Lomeli, Maria and Hambro, Eric and Zettlemoyer, Luke and Cancedda, Nicola and Scialom, Thomas},
  booktitle    = {Advances in Neural Information Processing Systems},
  year         = {2023}
}

@inproceedings{reflexion,
  title        = {Reflexion: Language Agents with Verbal Reinforcement Learning},
  author       = {Shinn, Noah and Cassano, Federico and Gopinath, Ashwin and Narasimhan, Karthik and Yao, Shunyu},
  booktitle    = {Advances in Neural Information Processing Systems},
  year         = {2023}
}

@inproceedings{camel,
  title        = {{CAMEL}: Communicative Agents for ``Mind'' Exploration of Large Language Model Society},
  author       = {Li, Guohao and Hammoud, Hasan Abed Al Kader and Itani, Hani and Khizbullin, Dmitrii and Ghanem, Bernard},
  booktitle    = {Advances in Neural Information Processing Systems},
  year         = {2023}
}

@inproceedings{autogen,
  title        = {{AutoGen}: Enabling Next-Gen {LLM} Applications via Multi-Agent Conversations},
  author       = {Wu, Qingyun and Bansal, Gagan and Zhang, Jieyu and Wu, Yiran and Li, Beibin and Zhu, Erkang and Jiang, Li and Zhang, Xiaoyun and Zhang, Shaokun and Liu, Jiale and Awadallah, Ahmed Hassan and White, Ryen W. and Burger, Doug and Wang, Chi},
  booktitle    = {First Conference on Language Modeling},
  year         = {2024}
}

@inproceedings{chatdev,
  title        = {{ChatDev}: Communicative Agents for Software Development},
  author       = {Qian, Chen and Liu, Wei and Liu, Hongzhang and Chen, Nuo and Dang, Yufan and Li, Jiahao and Yang, Cheng and Chen, Weize and Su, Yusheng and Cong, Xin and Xu, Juyuan and Li, Dahai and Liu, Zhiyuan and Sun, Maosong},
  booktitle    = {Proceedings of the 62nd Annual Meeting of the Association for Computational Linguistics},
  pages        = {15174-15186},
  year         = {2024},
  doi          = {10.18653/v1/2024.acl-long.810}
}

@inproceedings{metagpt,
  title        = {{MetaGPT}: Meta Programming for A Multi-Agent Collaborative Framework},
  author       = {Hong, Sirui and Zhuge, Mingchen and Chen, Jonathan and Zheng, Xiawu and Cheng, Yuheng and Wang, Jinlin and Zhang, Ceyao and Wang, Zili and Yau, Steven Ka Shing and Lin, Zijuan and Zhou, Liyang and Ran, Chenyu and Xiao, Lingfeng and Wu, Chenglin and Schmidhuber, J{\"u}rgen},
  booktitle    = {International Conference on Learning Representations},
  year         = {2024}
}

@inproceedings{llmcompiler,
  title        = {An {LLM} Compiler for Parallel Function Calling},
  author       = {Kim, Sehoon and Moon, Suhong and Tabrizi, Ryan and Lee, Nicholas and Mahoney, Michael W. and Keutzer, Kurt and Gholami, Amir},
  booktitle    = {International Conference on Machine Learning},
  year         = {2024}
}

@inproceedings{got,
  title        = {Graph of Thoughts: Solving Elaborate Problems with Large Language Models},
  author       = {Besta, Maciej and Blach, Nils and Kubicek, Ales and Gerstenberger, Robert and Podstawski, Michal and Gianinazzi, Lukas and Gajda, Joanna and Lehmann, Tomasz and Niewiadomski, Hubert and Nyczyk, Piotr and Hoefler, Torsten},
  booktitle    = {AAAI Conference on Artificial Intelligence},
  year         = {2024}
}

@inproceedings{dylan,
  title        = {A Dynamic {LLM}-Powered Agent Network for Task-Oriented Agent Collaboration},
  author       = {Liu, Zijun and Zhang, Yanzhe and Li, Peng and Liu, Yang and Yang, Diyi},
  booktitle    = {Conference on Language Modeling},
  year         = {2024}
}

@inproceedings{swebench,
  title        = {{SWE-bench}: Can Language Models Resolve Real-World {GitHub} Issues?},
  author       = {Jimenez, Carlos E. and Yang, John and Wettig, Alexander and Yao, Shunyu and Pei, Kexin and Press, Ofir and Narasimhan, Karthik},
  booktitle    = {International Conference on Learning Representations},
  year         = {2024}
}

@inproceedings{sweagent,
  title        = {{SWE-agent}: Agent-Computer Interfaces Enable Automated Software Engineering},
  author       = {Yang, John and Jimenez, Carlos E. and Wettig, Alexander and Lieret, Kilian and Yao, Shunyu and Narasimhan, Karthik and Press, Ofir},
  booktitle    = {Advances in Neural Information Processing Systems},
  year         = {2024}
}

@inproceedings{taubench,
  title        = {{$\tau$-bench}: A Benchmark for Tool-Agent-User Interaction in Real-World Domains},
  author       = {Yao, Shunyu and Shinn, Noah and Razavi, Pedram and Narasimhan, Karthik},
  booktitle    = {International Conference on Learning Representations},
  year         = {2025}
}

@inproceedings{agentbench,
  title        = {{AgentBench}: Evaluating {LLMs} as Agents},
  author       = {Liu, Xiao and Yu, Hao and Zhang, Hanchen and Xu, Yifan and Lei, Xuanyu and Lai, Hanyu and Gu, Yu and Ding, Hangliang and Men, Kaiwen and Yang, Kejuan and Zhang, Shudan and Deng, Xiang and Zeng, Aohan and Du, Zhengxiao and Zhang, Chenhui and Shen, Sheng and Zhang, Tianjun and Su, Yu and Sun, Huan and Huang, Minlie and Dong, Yuxiao and Tang, Jie},
  booktitle    = {International Conference on Learning Representations},
  year         = {2024}
}

@inproceedings{vaswani2017,
  title        = {Attention Is All You Need},
  author       = {Vaswani, Ashish and Shazeer, Noam and Parmar, Niki and Uszkoreit, Jakob and Jones, Llion and Gomez, Aidan N. and Kaiser, Lukasz and Polosukhin, Illia},
  booktitle    = {Advances in Neural Information Processing Systems},
  year         = {2017}
}

@inproceedings{vllm,
  title        = {Efficient Memory Management for Large Language Model Serving with {PagedAttention}},
  author       = {Kwon, Woosuk and Li, Zhuohan and Zhuang, Siyuan and Sheng, Ying and Zheng, Lianmin and Yu, Cody Hao and Gonzalez, Joseph E. and Zhang, Hao and Stoica, Ion},
  booktitle    = {ACM Symposium on Operating Systems Principles},
  year         = {2023}
}

@inproceedings{sglang,
  title        = {{SGLang}: Efficient Execution of Structured Language Model Programs},
  author       = {Zheng, Lianmin and Yin, Liangsheng and Xie, Zhiqiang and Sun, Chuyue and Huang, Jeff and Yu, Cody Hao and Cao, Shiyi and Kozyrakis, Christos and Stoica, Ion and Gonzalez, Joseph E. and Barrett, Clark W. and Sheng, Ying},
  booktitle    = {Advances in Neural Information Processing Systems},
  year         = {2024}
}

@inproceedings{kvflow,
  title        = {{KVFlow}: Efficient Prefix Caching for Accelerating {LLM}-Based Multi-Agent Workflows},
  author       = {Pan, Zaifeng and Patel, Ajjkumar and Shen, Yipeng and Hu, Zhengding and Guan, Yue and Li, Wan-Lu and Qin, Lianhui and Wang, Yida and Ding, Yufei},
  booktitle    = {Advances in Neural Information Processing Systems},
  year         = {2025}
}

@inproceedings{flashagents,
  title        = {{FlashAgents}: Accelerating Multi-Agent {LLM} Systems via Streaming Prefill Overlap},
  author       = {Fang, Taosong and Zheng, Zhen and Ma, Zhengzhao and Lu, Yaojie and Lin, Hongyu and Han, Xianpei and Sun, Le},
  booktitle    = {Proceedings of Machine Learning and Systems},
  year         = {2026}
}

@inproceedings{hotpotqa,
  title        = {{HotpotQA}: A Dataset for Diverse, Explainable Multi-hop Question Answering},
  author       = {Yang, Zhilin and Qi, Peng and Zhang, Saizheng and Bengio, Yoshua and Cohen, William W. and Salakhutdinov, Ruslan and Manning, Christopher D.},
  booktitle    = {Conference on Empirical Methods in Natural Language Processing},
  year         = {2018}
}

@article{musique,
  title        = {{MuSiQue}: Multihop Questions via Single-hop Question Composition},
  author       = {Trivedi, Harsh and Balasubramanian, Niranjan and Khot, Tushar and Sabharwal, Ashish},
  journal      = {Transactions of the Association for Computational Linguistics},
  volume       = {10},
  pages        = {539-554},
  year         = {2022}
}

@inproceedings{twowikimultihopqa,
  title        = {Constructing A Multi-hop {QA} Dataset for Comprehensive Evaluation of Reasoning Steps},
  author       = {Ho, Xanh and Duong Nguyen, Anh-Khoa and Sugawara, Saku and Aizawa, Akiko},
  booktitle    = {Proceedings of the 28th International Conference on Computational Linguistics},
  pages        = {6609-6625},
  year         = {2020},
  doi          = {10.18653/v1/2020.coling-main.580}
}

@inproceedings{rag,
  title        = {Retrieval-Augmented Generation for Knowledge-Intensive {NLP} Tasks},
  author       = {Lewis, Patrick and Perez, Ethan and Piktus, Aleksandra and Petroni, Fabio and Karpukhin, Vladimir and Goyal, Naman and K{\"u}ttler, Heinrich and Lewis, Mike and Yih, Wen-tau and Rockt{\"a}schel, Tim and Riedel, Sebastian and Kiela, Douwe},
  booktitle    = {Advances in Neural Information Processing Systems},
  pages        = {9459-9474},
  year         = {2020}
}

@inproceedings{cag,
  title        = {Don't Do {RAG}: When Cache-Augmented Generation is All You Need for Knowledge Tasks},
  author       = {Chan, Brian J. and Chen, Chao-Ting and Cheng, Jui-Hung and Huang, Hen-Hsen},
  booktitle    = {Companion Proceedings of the ACM on Web Conference 2025},
  pages        = {893-897},
  year         = {2025},
  doi          = {10.1145/3701716.3715490}
}

@inproceedings{lats,
  title        = {Language Agent Tree Search Unifies Reasoning, Acting, and Planning in Language Models},
  author       = {Zhou, Andy and Yan, Kai and Shlapentokh-Rothman, Michal and Wang, Haohan and Wang, Yu-Xiong},
  booktitle    = {International Conference on Machine Learning},
  year         = {2024}
}

@inproceedings{promptcache,
  title        = {Prompt Cache: Modular Attention Reuse for Low-Latency Inference},
  author       = {Gim, In and Chen, Guojun and Lee, Seung-seob and Sarda, Nikhil and Khandelwal, Anurag and Zhong, Lin},
  booktitle    = {Proceedings of Machine Learning and Systems},
  year         = {2024}
}

@inproceedings{cacheblend,
  title        = {{CacheBlend}: Fast Large Language Model Serving for {RAG} with Cached Knowledge Fusion},
  author       = {Yao, Jiayi and Li, Hanchen and Liu, Yuhan and Ray, Siddhant and Cheng, Yihua and Zhang, Qizheng and Du, Kuntai and Lu, Shan and Jiang, Junchen},
  booktitle    = {European Conference on Computer Systems},
  year         = {2025}
}

@inproceedings{flexgen,
  title        = {{FlexGen}: High-Throughput Generative Inference of Large Language Models with a Single {GPU}},
  author       = {Sheng, Ying and Zheng, Lianmin and Yuan, Binhang and Li, Zhuohan and Ryabinin, Max and Chen, Beidi and Liang, Percy and Re, Christopher and Stoica, Ion and Zhang, Ce},
  booktitle    = {International Conference on Machine Learning},
  pages        = {31094-31116},
  year         = {2023}
}

@inproceedings{llminflash,
  title        = {{LLM} in a flash: Efficient Large Language Model Inference with Limited Memory},
  author       = {Alizadeh, Keivan and Mirzadeh, Seyed Iman and Belenko, Dmitry and Khatamifard, S and Cho, Minsik and Del Mundo, Carlo C and Rastegari, Mohammad and Farajtabar, Mehrdad},
  booktitle    = {Proceedings of the 62nd Annual Meeting of the Association for Computational Linguistics},
  pages        = {12562-12584},
  year         = {2024},
  doi          = {10.18653/v1/2024.acl-long.678}
}

@inproceedings{cachegen,
  title        = {{CacheGen}: {KV} Cache Compression and Streaming for Fast Large Language Model Serving},
  author       = {Liu, Yuhan and Li, Hanchen and Cheng, Yihua and Ray, Siddhant and Huang, Yuyang and Zhang, Qizheng and Du, Kuntai and Yao, Jiayi and Lu, Shan and Ananthanarayanan, Ganesh and Maire, Michael and Hoffmann, Henry and Holtzman, Ari and Jiang, Junchen},
  booktitle    = {ACM SIGCOMM Conference},
  pages        = {38-56},
  year         = {2024},
  doi          = {10.1145/3651890.3672274}
}

@inproceedings{infinigen,
  title        = {{InfiniGen}: Efficient Generative Inference of Large Language Models with Dynamic {KV} Cache Management},
  author       = {Lee, Wonbeom and Lee, Jungi and Seo, Junghwan and Sim, Jaewoong},
  booktitle    = {USENIX Symposium on Operating Systems Design and Implementation},
  pages        = {155-172},
  year         = {2024}
}

@inproceedings{mooncake,
  title        = {{Mooncake}: Trading More Storage for Less Computation {\textemdash} {A} {KVCache-centric} Architecture for Serving {LLM} Chatbot},
  author       = {Qin, Ruoyu and Li, Zheming and He, Weiran and Cui, Jialei and Ren, Feng and Zhang, Mingxing and Wu, Yongwei and Zheng, Weimin and Xu, Xinran},
  booktitle    = {USENIX Conference on File and Storage Technologies},
  pages        = {155-170},
  year         = {2025}
}

@article{memserve,
  title        = {{MemServe}: Context Caching for Disaggregated {LLM} Serving with Elastic Memory Pool},
  author       = {Hu, Cunchen and Huang, Heyang and Hu, Junhao and Xu, Jiang and Chen, Xusheng and Xie, Tao and Wang, Chenxi and Wang, Sa and Bao, Yungang and Sun, Ninghui and Shan, Yizhou},
  journal      = {arXiv preprint arXiv:2406.17565},
  year         = {2024}
}

@article{lmcache,
  title        = {{LMCache}: An Efficient {KV} Cache Layer for Enterprise-Scale {LLM} Inference},
  author       = {Liu, Yuhan and Cheng, Yihua and Yao, Jiayi and An, Yuwei and Chen, Xiaokun and Feng, Shaoting and Huang, Yuyang and Shen, Samuel and Zhang, Rui and Du, Kuntai and Jiang, Junchen},
  journal      = {arXiv preprint arXiv:2510.09665},
  year         = {2025}
}

@INPROCEEDINGS{costofdynamic,
  author={Kim, Jiin and Shin, Byeongjun and Chung, Jinha and Rhu, Minsoo},
  booktitle={2026 IEEE International Symposium on High Performance Computer Architecture (HPCA)},
  title={The Cost of Dynamic Reasoning: Demystifying AI Agents and Test-Time Scaling from an AI Infrastructure Perspective},
  year={2026},
  volume={},
  number={},
  pages={1-16},
  doi={10.1109/HPCA68181.2026.11408569}}

@inproceedings{parrot,
  title        = {Parrot: Efficient Serving of {LLM}-based Applications with Semantic Variable},
  author       = {Lin, Chaofan and Han, Zhenhua and Zhang, Chengruidong and Yang, Yuqing and Yang, Fan and Chen, Chen and Qiu, Lili},
  booktitle    = {USENIX Symposium on Operating Systems Design and Implementation},
  year         = {2024}
}

@inproceedings{autellix,
  title        = {Agentix: An Efficient Serving Engine for {LLM} Agents as General Programs},
  author       = {Luo, Michael and Shi, Xiaoxiang and Cai, Colin and Zhang, Tianjun and Wong, Justin and Wang, Yichuan and Wang, Chi and Huang, Yanping and Chen, Zhifeng and Gonzalez, Joseph E. and Stoica, Ion},
  booktitle    = {USENIX Symposium on Networked Systems Design and Implementation},
  year         = {2026},
  note         = {Previously circulated as ``Autellix'' (arXiv:2502.13965)}
}

@article{tokencake,
  title        = {{TokenCake}: A {KV}-Cache-centric Serving Framework for {LLM}-based Multi-Agent Applications},
  author       = {Bian, Zhuohang and Wu, Feiyang and Li, Zhuoran and Ma, Teng and Zhuo, Youwei},
  journal      = {arXiv preprint arXiv:2510.18586},
  year         = {2025}
}

@article{paste,
  title        = {Parallelizing Tool Execution and {LLM} Generation for Low-Latency Agent Serving},
  author       = {Sui, Yifan and Zhao, Han and Ma, Rui and He, Zhiyuan and Wang, Hao and Li, Jianxun and Yang, Yuqing},
  journal      = {arXiv preprint arXiv:2603.18897},
  year         = {2026}
}

@inproceedings{gptcache,
  title        = {{GPTCache}: An Open-Source Semantic Cache for {LLM} Applications Enabling Faster Answers and Cost Savings},
  author       = {Fu Bang},
  booktitle    = {Proceedings of the 3rd Workshop for Natural Language Processing Open Source Software (NLP-OSS 2023)},
  pages        = {212-218},
  year         = {2023},
  doi          = {10.18653/v1/2023.nlposs-1.24}
}

@inproceedings{abmcts,
  title        = {Wider or Deeper? Scaling {LLM} Inference-Time Compute with Adaptive Branching Tree Search},
  author       = {Inoue, Yuichi and Misaki, Kou and Imajuku, Yuki and Kuroki, So and Nakamura, Taishi and Akiba, Takuya},
  booktitle    = {Advances in Neural Information Processing Systems},
  year         = {2025}
}

@inproceedings{aflow,
  title        = {{AFlow}: Automating Agentic Workflow Generation},
  author       = {Zhang, Jiayi and Xiang, Jinyu and Yu, Zhaoyang and Teng, Fengwei and Chen, XiongHui and Chen, Jiaqi and Zhuge, Mingchen and Cheng, Xin and Hong, Sirui and Wang, Jinlin and Zheng, Bingnan and Liu, Bang and Luo, Yuyu and Wu, Chenglin},
  booktitle    = {International Conference on Learning Representations},
  year         = {2025}
}

@inproceedings{rstarmath,
  title        = {{rStar-Math}: Small {LLMs} Can Master Math Reasoning with Self-Evolved Deep Thinking},
  author       = {Guan, Xinyu and Zhang, Li Lyna and Liu, Yifei and Shang, Ning and Sun, Youran and Zhu, Yi and Yang, Fan and Yang, Mao},
  booktitle    = {International Conference on Machine Learning},
  year         = {2025}
}

@inproceedings{swesearch,
  title        = {{SWE-Search}: Enhancing Software Agents with Monte Carlo Tree Search and Iterative Refinement},
  author       = {Antoniades, Antonis and {\"O}rwall, Albert and Zhang, Kexun and Xie, Yuxi and Goyal, Anirudh and Wang, William},
  booktitle    = {International Conference on Learning Representations},
  year         = {2025}
}

@inproceedings{exact,
  title        = {{ExACT}: Teaching {AI} Agents to Explore with Reflective-{MCTS} and Exploratory Learning},
  author       = {Yu, Xiao and Peng, Baolin and Vajipey, Vineeth and Cheng, Hao and Galley, Michel and Gao, Jianfeng and Yu, Zhou},
  booktitle    = {International Conference on Learning Representations},
  year         = {2025}
}

@inproceedings{mctsrag,
  title        = {{MCTS-RAG}: Enhancing Retrieval-Augmented Generation with Monte Carlo Tree Search},
  author       = {Hu, Yunhai and Zhao, Yilun and Zhao, Chen and Cohan, Arman},
  booktitle      = {Findings of the Association for Computational Linguistics: EMNLP},
  year         = {2025},
  pages = {12581-12597},
  doi = {10.18653/v1/2025.findings-emnlp.672}
}

@misc{llama4,
  title        = {The {Llama 4} Herd: The Beginning of a New Era of Natively Multimodal {AI} Innovation},
  author       = {{Meta AI}},
  year         = {2025},
  howpublished = {\url{https://ai.meta.com/blog/llama-4-multimodal-intelligence/}},
  note         = {Llama 4 Maverick: 400B total / 17B active (MoE)}
}

@article{deepseekv3,
  title        = {{DeepSeek-V3} Technical Report},
  author       = {{DeepSeek-AI}},
  journal      = {arXiv preprint arXiv:2412.19437},
  year         = {2024}
}

@inproceedings{math,
  title={Measuring Mathematical Problem Solving With the MATH Dataset},
  author={Dan Hendrycks and Collin Burns and Saurav Kadavath and Akul Arora and Steven Basart and Eric Tang and Dawn Song and Jacob Steinhardt},
  booktitle={Proceedings of the Neural Information Processing Systems Track on Datasets and Benchmarks},
  year={2021}
}

@article{humaneval,
  title={Evaluating Large Language Models Trained on Code},
  author={Mark Chen and Jerry Tworek and Heewoo Jun and Qiming Yuan and Henrique Ponde de Oliveira Pinto and Jared Kaplan and Harri Edwards and Yuri Burda and Nicholas Joseph and Greg Brockman and Alex Ray and Raul Puri and Gretchen Krueger and Michael Petrov and Heidy Khlaaf and Girish Sastry and Pamela Mishkin and Brooke Chan and Scott Gray and Nick Ryder and Mikhail Pavlov and Alethea Power and Lukasz Kaiser and Mohammad Bavarian and Clemens Winter and Philippe Tillet and Felipe Petroski Such and Dave Cummings and Matthias Plappert and Fotios Chantzis and Elizabeth Barnes and Ariel Herbert-Voss and William Hebgen Guss and Alex Nichol and Alex Paino and Nikolas Tezak and Jie Tang and Igor Babuschkin and Suchir Balaji and Shantanu Jain and William Saunders and Christopher Hesse and Andrew N. Carr and Jan Leike and Josh Achiam and Vedant Misra and Evan Morikawa and Alec Radford and Matthew Knight and Miles Brundage and Mira Murati and Katie Mayer and Peter Welinder and Bob McGrew and Dario Amodei and Sam McCandlish and Ilya Sutskever and Wojciech Zaremba},
  year={2021},
  journal={arXiv preprint arXiv:2107.03374},
  eprint={2107.03374},
  archivePrefix={arXiv},
  primaryClass={cs.LG}
}

@misc{kimik3,
      title={{Kimi K3}: Open Frontier Intelligence},
      author={{Kimi Team} and Tongtong Bai and Yifan Bai and Yiping Bao and M. C. and Jianfeng Cai and Xinyuan Cai and Peizhou Cao and Yuxuan Cao and Ziwei Chai and Y. Charles and H. S. Che and Guanduo Chen and Guangyu Chen and Guanzheng Chen and Huarong Chen and Jia Chen and Jianlong Chen and Jun Chen and Kexin Chen and Peng Chen and Ruijue Chen and Wentao Chen and Xin Chen and Yang Chen and Yanru Chen and Yifei Chen and Yingjiang Chen and Yuankun Chen and Yujie Chen and Yutian Chen and Zhirong Chen and Dazhi Cheng and Yean Cheng and Jialei Cui and Jingbing Cui and Anqi Dai and Jiaqi Deng and Hao Ding and Rui Ding and Shaofeng Ding and Mengfan Dong and Mengnan Dong and Yuhao Dong and Yuxin Dong and Angang Du and Chenzhuang Du and Dikang Du and Jusen Du and Yulun Du and Yu Fan and Jing Feng and Qiulin Feng and Yichen Feng and Kelin Fu and Qiang Fu and Fuxuan Gao and Hongcheng Gao and Jingyue Gao and Tong Gao and Weijia Gao and Shangyi Geng and Jie Gong and Linhu Gong and Shengao Gong and Xiaochen Gong and Qizheng Gu and Yicheng Gu and Shuhao Guan and Haiqing Guo and Shiqi Guo and Xiang Guo and Zhengyan Guo and Beixi Hao and Wenxin Hao and Xiaoru Hao and Dailan He and Haotian He and Lehan He and Qi He and Weiran He and Xinran He and Xinyi He and Yibo He and Yunjia He and Chao Hong and Tiange Hong and Hao Hu and Jiaxi Hu and Ruikun Hu and Weiming Hu and Yangyang Hu and Zhenxing Hu and Liang Hua and Jinbin Huang and Ke Huang and Ruiyuan Huang and Siying Huang and Weixiao Huang and Yan Huang and Zhengjie Huang and Zhiqi Huang and Yulong Hui and Chaobo Jia and Yutong Jiang and Zhejun Jiang and Zuoyou Jiang and Wenyi Jin and Xinyi Jin and Yu Jing and Huanjun Kong and Guokun Lai and Aidi Li and Cheng Li and Chengyuan Li and Cong Li and Fang Li and Guanyu Li and Haoyang Li and Jia Li and Junxiong Li and Lei Li and Letian Li and Lincan Li and Weihong Li and Wentao Li and Xintong Li and Yang Li and Yishen Li and Yiwei Li and Yuxiao Li and Zhaowei Li and Zhaoxi Li and Zheming Li and Zhengxiao Li and Zhiyuan Li and Jiawei Lin and Xiaohan Lin and Yibo Lin and Zichao Lin and Ziyan Lin and Bill Liu and Boxiao Liu and Chuan Liu and Liang Liu and Shaowei Liu and Shudong Liu and Shuran Liu and Tianwei Liu and Weizhou Liu and Yangyang Liu and Yanming Liu and Yibo Liu and Yipeng Liu and Zhengying Liu and Zhiheng Liu and Enzhe Lu and Haoyu Lu and Linqiang Lu and Tingzhan Lu and Zhiyuan Lu and Aotian Luo and G. Luo and Junyu Luo and Yifan Luo and B. Lyu and Wenzhou Lyu and Shaoguang Mao and Yuan Mei and Xin Men and Minqing Ni and Yixuan Niu and Siyuan Pan and Shujun Peng and Zhangyang Qi and Ruoyu Qin and ZeChao Qin and Zeyu Qin and Haiquan Qiu and Jianxin Qiu and Jiezhong Qiu and Bowen Qu and Yuhao Qu and Zeyu Shang and Youbo Shao and Han Shen and Jincheng Shi and Juanfeng Shi and Lidong Shi and Shengyuan Shi and Wingchun Siu and Pengwei Song and Xiaoxi Song and Jianlin Su and Yunfeng Su and Zhaochen Su and Lin Sui and Jingsong Sun and Junyao Sun and Shaoning Sun and Shuzhe Sun and Tongyu Sun and Yujun Sun and Yunpeng Tai and Chuning Tang and Heyi Tang and Sirui Tang and Zecheng Tang and Chaoran Tian and Rongpeng Tian and Yu Tian and Wei Tu and Chensi Wang and Chuang Wang and Chunjie Wang and Dinglu Wang and Feng Wang and Hailong Wang and Haiming Wang and Hao Wang and Hao Wang and Huaqing Wang and Hui Wang and Jiayi Wang and Jinglong Wang and Jinhong Wang and Jiuzheng Wang and Linian Wang and Shaobo Wang and Shenzhi Wang and Shuyi Wang and Si Wang and Siyuan Wang and Tianfu Wang and Wenjue Wang and Xingran Wang and Xinmei Wang and Xinyuan Wang and Xusheng Wang and Yalin Wang and Yangkun Wang and Yao Wang and Yaoyu Wang and Yejie Wang and Yiqin Wang and Yucheng Wang and Yuzhi Wang and Zhaoji Wang and Zhaowei Wang and Zhengtao Wang and Zhenhao Wang and Zhongsheng Wang and Zifan Wang and Chu Wei and Ming Wei and Shouxin Wei and Zichen Wen and Fan Wu and Haoning Wu and Rucong Wu and Wenhao Wu and Xiaoxue Wu and Yingcong Wu and Yongqi Wu and Yuxin Wu and Zijian Wu and Xinglang Xian and Chenxuan Xiang and Yuye Xiang and Bocheng Xiao and Chenjun Xiao and Xin Xiao and Jin Xie and Xiaotong Xie and Yifeng Xie and Zhe Xie and Bowei Xing and Yiming Xiong and Baosheng Xu and Boyu Xu and Jiale Xu and Jianfan Xu and Jing Xu and Jinjing Xu and L. H. Xu and Qingtao Xu and Shuyao Xu and Suting Xu and Tiantian Xu and Tianxiang Xu and Weixin Xu and Xinran Xu and Yangchuan Xu and Ye Xu and Yueni Xu and Ziyao Xu and Haonan Xue and Junjie Yan and Yaoyao Yan and Fan Yang and Guangyao Yang and Hao Yang and Junwei Yang and Ruoyu Yang and Wenjie Yang and Xiaofei Yang and Xinyu Yang and Yi Yang and Yiling Yang and Ying Yang and Yuchen Yang and Zhen Yang and Zhilin Yang and Zian Yang and Zuhao Yang and Haotian Yao and Dan Ye and Haoran Ye and Wenjie Ye and Zhanbo Ye and Bohong Yin and Haoxiang Yin and Xietong Yin and Chengzhen Yu and Haozhen Yu and Longhui Yu and Shengnan Yu and Shuying Yu and Tianxiang Yu and Enming Yuan and Mengjie Yuan and Tongtian Yue and Wei Yue and Yang Yue and Dunyuan Zha and Haobing Zhan and B. H. Zhang and Dehao Zhang and Fei Zhang and Hao Zhang and Haoyuan Zhang and Huanyu Zhang and Jiapei Zhang and Jiaxuan Zhang and Jin Zhang and Kaiyi Zhang and Miaozhen Zhang and Puqi Zhang and Qinglei Zhang and Rong Zhang and Rui Zhang and Shaoshuai Zhang and Shiyi Zhang and Xiaobin Zhang and Xiaoyun Zhang and Y. Zhang and Yangkun Zhang and Ye Zhang and Yichi Zhang and Yikun Zhang and Yizhi Zhang and Yongting Zhang and Yu Zhang and Yutao Zhang and Yutong Zhang and Zheng Zhang and Zijing Zhang and Bin Zhao and Chenguang Zhao and Feifan Zhao and Jinglun Zhao and Jinxiang Zhao and Shuai Zhao and Wenshuo Zhao and Xiangyu Zhao and Xuanle Zhao and Yikai Zhao and Zijia Zhao and Haozhi Zheng and Huabin Zheng and Ruihan Zheng and Shaojie Zheng and Tengyang Zheng and Haofeng Zhong and Lei Zhong and Longguang Zhong and M. Zhou and Qiankang Zhou and Runjie Zhou and Ruozhang Zhou and Xinyu Zhou and Yiqiao Zhou and Zaida Zhou and Jinguo Zhu and Liya Zhu and Xinhao Zhu and Yangjunfeng Zhu and Yuxuan Zhu and Zhen Zhu and Chen Zhuang and Weiyu Zhuang and Xinxing Zu},
      year={2026},
      eprint={2607.24653},
      archivePrefix={arXiv},
      primaryClass={cs.CL},
      url={https://arxiv.org/abs/2607.24653},
}

@misc{llama3,
      title={The {Llama 3} Herd of Models},
      author={Aaron Grattafiori and Abhimanyu Dubey and Abhinav Jauhri and Abhinav Pandey and Abhishek Kadian and Ahmad Al-Dahle and Aiesha Letman and Akhil Mathur and Alan Schelten and Alex Vaughan and Amy Yang and Angela Fan and Anirudh Goyal and Anthony Hartshorn and Aobo Yang and Archi Mitra and Archie Sravankumar and Artem Korenev and Arthur Hinsvark and Arun Rao and Aston Zhang and Aurelien Rodriguez and Austen Gregerson and Ava Spataru and Baptiste Roziere and Bethany Biron and Binh Tang and Bobbie Chern and Charlotte Caucheteux and Chaya Nayak and Chloe Bi and Chris Marra and Chris McConnell and Christian Keller and Christophe Touret and Chunyang Wu and Corinne Wong and Cristian Canton Ferrer and Cyrus Nikolaidis and Damien Allonsius and Daniel Song and Danielle Pintz and Danny Livshits and Danny Wyatt and David Esiobu and Dhruv Choudhary and Dhruv Mahajan and Diego Garcia-Olano and Diego Perino and Dieuwke Hupkes and Egor Lakomkin and Ehab AlBadawy and Elina Lobanova and Emily Dinan and Eric Michael Smith and Filip Radenovic and Francisco Guzmán and Frank Zhang and Gabriel Synnaeve and Gabrielle Lee and Georgia Lewis Anderson and Govind Thattai and Graeme Nail and Gregoire Mialon and Guan Pang and Guillem Cucurell and Hailey Nguyen and Hannah Korevaar and Hu Xu and Hugo Touvron and Iliyan Zarov and Imanol Arrieta Ibarra and Isabel Kloumann and Ishan Misra and Ivan Evtimov and Jack Zhang and Jade Copet and Jaewon Lee and Jan Geffert and Jana Vranes and Jason Park and Jay Mahadeokar and Jeet Shah and Jelmer van der Linde and Jennifer Billock and Jenny Hong and Jenya Lee and Jeremy Fu and Jianfeng Chi and Jianyu Huang and Jiawen Liu and Jie Wang and Jiecao Yu and Joanna Bitton and Joe Spisak and Jongsoo Park and Joseph Rocca and Joshua Johnstun and Joshua Saxe and Junteng Jia and Kalyan Vasuden Alwala and Karthik Prasad and Kartikeya Upasani and Kate Plawiak and Ke Li and Kenneth Heafield and Kevin Stone and Khalid El-Arini and Krithika Iyer and Kshitiz Malik and Kuenley Chiu and Kunal Bhalla and Kushal Lakhotia and Lauren Rantala-Yeary and Laurens van der Maaten and Lawrence Chen and Liang Tan and Liz Jenkins and Louis Martin and Lovish Madaan and Lubo Malo and Lukas Blecher and Lukas Landzaat and Luke de Oliveira and Madeline Muzzi and Mahesh Pasupuleti and Mannat Singh and Manohar Paluri and Marcin Kardas and Maria Tsimpoukelli and Mathew Oldham and Mathieu Rita and Maya Pavlova and Melanie Kambadur and Mike Lewis and Min Si and Mitesh Kumar Singh and Mona Hassan and Naman Goyal and Narjes Torabi and Nikolay Bashlykov and Nikolay Bogoychev and Niladri Chatterji and Ning Zhang and Olivier Duchenne and Onur Çelebi and Patrick Alrassy and Pengchuan Zhang and Pengwei Li and Petar Vasic and Peter Weng and Prajjwal Bhargava and Pratik Dubal and Praveen Krishnan and Punit Singh Koura and Puxin Xu and Qing He and Qingxiao Dong and Ragavan Srinivasan and Raj Ganapathy and Ramon Calderer and Ricardo Silveira Cabral and Robert Stojnic and Roberta Raileanu and Rohan Maheswari and Rohit Girdhar and Rohit Patel and Romain Sauvestre and Ronnie Polidoro and Roshan Sumbaly and Ross Taylor and Ruan Silva and Rui Hou and Rui Wang and Saghar Hosseini and Sahana Chennabasappa and Sanjay Singh and Sean Bell and Seohyun Sonia Kim and Sergey Edunov and Shaoliang Nie and Sharan Narang and Sharath Raparthy and Sheng Shen and Shengye Wan and Shruti Bhosale and Shun Zhang and Simon Vandenhende and Soumya Batra and Spencer Whitman and Sten Sootla and Stephane Collot and Suchin Gururangan and Sydney Borodinsky and Tamar Herman and Tara Fowler and Tarek Sheasha and Thomas Georgiou and Thomas Scialom and Tobias Speckbacher and Todor Mihaylov and Tong Xiao and Ujjwal Karn and Vedanuj Goswami and Vibhor Gupta and Vignesh Ramanathan and Viktor Kerkez and Vincent Gonguet and Virginie Do and Vish Vogeti and Vítor Albiero and Vladan Petrovic and Weiwei Chu and Wenhan Xiong and Wenyin Fu and Whitney Meers and Xavier Martinet and Xiaodong Wang and Xiaofang Wang and Xiaoqing Ellen Tan and Xide Xia and Xinfeng Xie and Xuchao Jia and Xuewei Wang and Yaelle Goldschlag and Yashesh Gaur and Yasmine Babaei and Yi Wen and Yiwen Song and Yuchen Zhang and Yue Li and Yuning Mao and Zacharie Delpierre Coudert and Zheng Yan and Zhengxing Chen and Zoe Papakipos and Aaditya Singh and Aayushi Srivastava and Abha Jain and Adam Kelsey and Adam Shajnfeld and Adithya Gangidi and Adolfo Victoria and Ahuva Goldstand and Ajay Menon and Ajay Sharma and Alex Boesenberg and Alexei Baevski and Allie Feinstein and Amanda Kallet and Amit Sangani and Amos Teo and Anam Yunus and Andrei Lupu and Andres Alvarado and Andrew Caples and Andrew Gu and Andrew Ho and Andrew Poulton and Andrew Ryan and Ankit Ramchandani and Annie Dong and Annie Franco and Anuj Goyal and Aparajita Saraf and Arkabandhu Chowdhury and Ashley Gabriel and Ashwin Bharambe and Assaf Eisenman and Azadeh Yazdan and Beau James and Ben Maurer and Benjamin Leonhardi and Bernie Huang and Beth Loyd and Beto De Paola and Bhargavi Paranjape and Bing Liu and Bo Wu and Boyu Ni and Braden Hancock and Bram Wasti and Brandon Spence and Brani Stojkovic and Brian Gamido and Britt Montalvo and Carl Parker and Carly Burton and Catalina Mejia and Ce Liu and Changhan Wang and Changkyu Kim and Chao Zhou and Chester Hu and Ching-Hsiang Chu and Chris Cai and Chris Tindal and Christoph Feichtenhofer and Cynthia Gao and Damon Civin and Dana Beaty and Daniel Kreymer and Daniel Li and David Adkins and David Xu and Davide Testuggine and Delia David and Devi Parikh and Diana Liskovich and Didem Foss and Dingkang Wang and Duc Le and Dustin Holland and Edward Dowling and Eissa Jamil and Elaine Montgomery and Eleonora Presani and Emily Hahn and Emily Wood and Eric-Tuan Le and Erik Brinkman and Esteban Arcaute and Evan Dunbar and Evan Smothers and Fei Sun and Felix Kreuk and Feng Tian and Filippos Kokkinos and Firat Ozgenel and Francesco Caggioni and Frank Kanayet and Frank Seide and Gabriela Medina Florez and Gabriella Schwarz and Gada Badeer and Georgia Swee and Gil Halpern and Grant Herman and Grigory Sizov and Guangyi Zhang and Guna Lakshminarayanan and Hakan Inan and Hamid Shojanazeri and Han Zou and Hannah Wang and Hanwen Zha and Haroun Habeeb and Harrison Rudolph and Helen Suk and Henry Aspegren and Hunter Goldman and Hongyuan Zhan and Ibrahim Damlaj and Igor Molybog and Igor Tufanov and Ilias Leontiadis and Irina-Elena Veliche and Itai Gat and Jake Weissman and James Geboski and James Kohli and Janice Lam and Japhet Asher and Jean-Baptiste Gaya and Jeff Marcus and Jeff Tang and Jennifer Chan and Jenny Zhen and Jeremy Reizenstein and Jeremy Teboul and Jessica Zhong and Jian Jin and Jingyi Yang and Joe Cummings and Jon Carvill and Jon Shepard and Jonathan McPhie and Jonathan Torres and Josh Ginsburg and Junjie Wang and Kai Wu and Kam Hou U and Karan Saxena and Kartikay Khandelwal and Katayoun Zand and Kathy Matosich and Kaushik Veeraraghavan and Kelly Michelena and Keqian Li and Kiran Jagadeesh and Kun Huang and Kunal Chawla and Kyle Huang and Lailin Chen and Lakshya Garg and Lavender A and Leandro Silva and Lee Bell and Lei Zhang and Liangpeng Guo and Licheng Yu and Liron Moshkovich and Luca Wehrstedt and Madian Khabsa and Manav Avalani and Manish Bhatt and Martynas Mankus and Matan Hasson and Matthew Lennie and Matthias Reso and Maxim Groshev and Maxim Naumov and Maya Lathi and Meghan Keneally and Miao Liu and Michael L. Seltzer and Michal Valko and Michelle Restrepo and Mihir Patel and Mik Vyatskov and Mikayel Samvelyan and Mike Clark and Mike Macey and Mike Wang and Miquel Jubert Hermoso and Mo Metanat and Mohammad Rastegari and Munish Bansal and Nandhini Santhanam and Natascha Parks and Natasha White and Navyata Bawa and Nayan Singhal and Nick Egebo and Nicolas Usunier and Nikhil Mehta and Nikolay Pavlovich Laptev and Ning Dong and Norman Cheng and Oleg Chernoguz and Olivia Hart and Omkar Salpekar and Ozlem Kalinli and Parkin Kent and Parth Parekh and Paul Saab and Pavan Balaji and Pedro Rittner and Philip Bontrager and Pierre Roux and Piotr Dollar and Polina Zvyagina and Prashant Ratanchandani and Pritish Yuvraj and Qian Liang and Rachad Alao and Rachel Rodriguez and Rafi Ayub and Raghotham Murthy and Raghu Nayani and Rahul Mitra and Rangaprabhu Parthasarathy and Raymond Li and Rebekkah Hogan and Robin Battey and Rocky Wang and Russ Howes and Ruty Rinott and Sachin Mehta and Sachin Siby and Sai Jayesh Bondu and Samyak Datta and Sara Chugh and Sara Hunt and Sargun Dhillon and Sasha Sidorov and Satadru Pan and Saurabh Mahajan and Saurabh Verma and Seiji Yamamoto and Sharadh Ramaswamy and Shaun Lindsay and Shaun Lindsay and Sheng Feng and Shenghao Lin and Shengxin Cindy Zha and Shishir Patil and Shiva Shankar and Shuqiang Zhang and Shuqiang Zhang and Sinong Wang and Sneha Agarwal and Soji Sajuyigbe and Soumith Chintala and Stephanie Max and Stephen Chen and Steve Kehoe and Steve Satterfield and Sudarshan Govindaprasad and Sumit Gupta and Summer Deng and Sungmin Cho and Sunny Virk and Suraj Subramanian and Sy Choudhury and Sydney Goldman and Tal Remez and Tamar Glaser and Tamara Best and Thilo Koehler and Thomas Robinson and Tianhe Li and Tianjun Zhang and Tim Matthews and Timothy Chou and Tzook Shaked and Varun Vontimitta and Victoria Ajayi and Victoria Montanez and Vijai Mohan and Vinay Satish Kumar and Vishal Mangla and Vlad Ionescu and Vlad Poenaru and Vlad Tiberiu Mihailescu and Vladimir Ivanov and Wei Li and Wenchen Wang and Wenwen Jiang and Wes Bouaziz and Will Constable and Xiaocheng Tang and Xiaojian Wu and Xiaolan Wang and Xilun Wu and Xinbo Gao and Yaniv Kleinman and Yanjun Chen and Ye Hu and Ye Jia and Ye Qi and Yenda Li and Yilin Zhang and Ying Zhang and Yossi Adi and Youngjin Nam and Yu Wang and Yu Zhao and Yuchen Hao and Yundi Qian and Yunlu Li and Yuzi He and Zach Rait and Zachary DeVito and Zef Rosnbrick and Zhaoduo Wen and Zhenyu Yang and Zhiwei Zhao and Zhiyu Ma},
      year={2024},
      eprint={2407.21783},
      archivePrefix={arXiv},
      primaryClass={cs.AI},
      url={https://arxiv.org/abs/2407.21783},
}

@inproceedings{venti,
  title        = {{Venti}: A New Approach to Archival Storage},
  author       = {Quinlan, Sean and Dorward, Sean},
  booktitle    = {USENIX Conference on File and Storage Technologies},
  pages        = {89-102},
  year         = {2002}
}

@inproceedings{lbfs,
  title        = {A Low-Bandwidth Network File System},
  author       = {Muthitacharoen, Athicha and Chen, Benjie and Mazi{\`e}res, David},
  booktitle    = {{ACM} Symposium on Operating Systems Principles},
  pages        = {174-187},
  year         = {2001}
}

@inproceedings{datadomain,
  title        = {Avoiding the Disk Bottleneck in the {Data Domain} Deduplication File System},
  author       = {Zhu, Benjamin and Li, Kai and Patterson, Hugo},
  booktitle    = {USENIX Conference on File and Storage Technologies},
  pages        = {269-282},
  year         = {2008}
}

@inproceedings{nix,
  title        = {{Nix}: A Safe and Policy-Free System for Software Deployment},
  author       = {Dolstra, Eelco and de Jonge, Merijn and Visser, Eelco},
  booktitle    = {Large Installation System Administration Conference},
  pages        = {79-92},
  year         = {2004}
}

@inproceedings{nectar,
  title        = {{Nectar}: Automatic Management of Data and Computation in Datacenters},
  author       = {Gunda, Pradeep Kumar and Ravindranath, Lenin and Thekkath, Chandramohan A. and Yu, Yuan and Zhuang, Li},
  booktitle    = {USENIX Symposium on Operating Systems Design and Implementation},
  pages        = {75-88},
  year         = {2010}
}

@article{alacarte,
  title        = {Build Systems {\`a} la Carte},
  author       = {Mokhov, Andrey and Mitchell, Neil and Peyton Jones, Simon},
  journal      = {Proceedings of the {ACM} on Programming Languages},
  volume       = {2},
  number       = {ICFP},
  pages        = {79:1-79:29},
  year         = {2018}
}

@article{michie,
  title        = {`{Memo}' Functions and Machine Learning},
  author       = {Michie, Donald},
  journal      = {Nature},
  volume       = {218},
  number       = {5136},
  pages        = {19-22},
  year         = {1968}
}

@article{btrfs,
  title        = {{BTRFS}: The {Linux} {B}-Tree Filesystem},
  author       = {Rodeh, Ohad and Bacik, Josef and Mason, Chris},
  journal      = {{ACM} Transactions on Storage},
  volume       = {9},
  number       = {3},
  pages        = {9:1-9:32},
  year         = {2013}
}

@inproceedings{tpp,
  title        = {{TPP}: Transparent Page Placement for {CXL}-Enabled Tiered-Memory},
  author       = {Maruf, Hasan Al and Wang, Hao and Dhanotia, Abhishek and Weiner, Johannes and Agarwal, Niket and Bhattacharya, Pallab and Petersen, Chris and Chowdhury, Mosharaf and Kanaujia, Shobhit and Chauhan, Prakash},
  booktitle    = {{ACM} International Conference on Architectural Support for Programming Languages and Operating Systems},
  pages        = {742-755},
  year         = {2023}
}

@inproceedings{cloudbuild,
  title        = {{CloudBuild}: {Microsoft's} Distributed and Caching Build Service},
  author       = {Esfahani, Hamed and Fietz, Jonas and Ke, Qi and Kolomiets, Alexei and Lan, Erica and Mavrinac, Erik and Schulte, Wolfram and Sanches, Newton and Kandula, Srikanth},
  booktitle    = {International Conference on Software Engineering Companion},
  pages        = {11-20},
  year         = {2016}
}

\end{document}